\documentclass[sigconf,pbalance]{acmart}

\usepackage{subcaption}
\usepackage{soul}
\usepackage{xspace}
\usepackage{xcolor}
\newcommand{\nickname}{3DALL-E\xspace}
\newcommand{\gdnickname}{GD\xspace}
\definecolor{lightgray}{gray}{0.6}

\newcommand{\revAdd}[1]{#1} % keep everything in the \revAdd{...} command (argument #1), no special formatting
\newcommand{\revDel}[1]{} % remove everything in the \revDel{...} command

\AtBeginDocument{%
  \providecommand\BibTeX{{%
    \normalfont B\kern-0.5em{\scshape i\kern-0.25em b}\kern-0.8em\TeX}}}

\copyrightyear{2023}
\acmYear{2023}
\setcopyright{acmlicensed}\acmConference[DIS '23]{Designing Interactive Systems Conference}{July 10--14, 2023}{Pittsburgh, PA, USA}
\acmBooktitle{Designing Interactive Systems Conference (DIS '23), July 10--14, 2023, Pittsburgh, PA, USA}
\acmPrice{15.00}
\acmDOI{10.1145/3563657.3596098}
\acmISBN{978-1-4503-9893-0/23/07}

\begin{document}

\title{\nickname: Integrating Text-to-Image AI in 3D Design Workflows}


\author{Vivian Liu}
\email{vivian@cs.columbia.edu}
\additionalaffiliation{%
  \institution{Columbia University}
  \city{New York}
  \state{New York}
  \country{USA}
}
\affiliation{%
  \institution{Autodesk Research}
  \city{Toronto}
  \state{Ontario}
  \country{Canada}
}

\author{Jo Vermeulen}
\email{jo.vermeulen@autodesk.com}
\affiliation{%
  \institution{Autodesk Research}
  \city{Toronto}
  \state{Ontario}
  \country{Canada}
}

\author{George Fitzmaurice}
\email{george.fitzmaurice@autodesk.com}
\affiliation{%
  \institution{Autodesk Research}
  \city{Toronto}
  \state{Ontario}
  \country{Canada}
}

\author{Justin Matejka}
\email{justin.matejka@autodesk.com}
\affiliation{%
  \institution{Autodesk Research}
  \city{Toronto}
  \state{Ontario}
  \country{Canada}
}

%\renewcommand{\shortauthors}{Liu, Vermeulen, Fitzmaurice, Matejka}

\begin{abstract}
Text-to-image AI are capable of generating novel images for inspiration, but their applications for 3D design workflows and how designers can build 3D models using AI-provided inspiration have not yet been explored. To investigate this, we integrated DALL-E, GPT-3, and CLIP within a CAD software in \textit{3DALL-E}, a plugin that 
 generates 2D image inspiration for 3D design. 3DALL-E allows users to construct text and image prompts based on what they are modeling. In a study with 13 designers, we found that designers saw great potential in 3DALL-E within their workflows and could use text-to-image AI to produce reference images, prevent design fixation, and inspire design considerations. We elaborate on prompting patterns observed across 3D modeling tasks and provide measures of prompt complexity observed across participants. From our findings, we discuss how 3DALL-E can merge with existing generative design workflows and propose prompt bibliographies as a form of human-AI design history.

\end{abstract}



% \ccsdesc[500]{Human-centered computing~Empirical studies in HCI}
% \ccsdesc[300]{Information systems~Language models}
% \ccsdesc[300]{Human-centered computing~Natural language interfaces}


\begin{teaserfigure}
  \includegraphics[width=\textwidth]{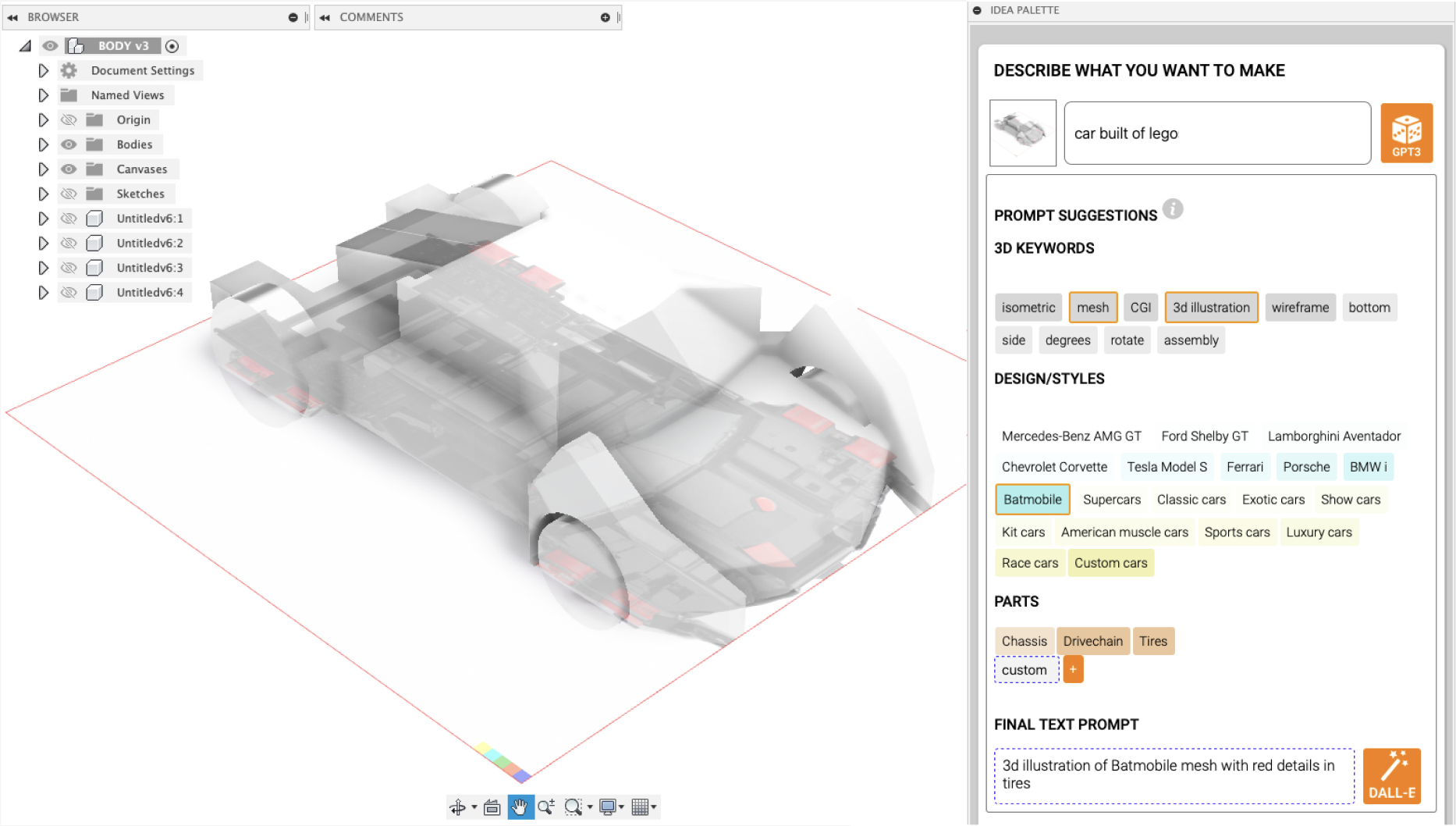}
  \caption{\nickname integrates a state-of-the-art text-to-image AI (DALL-E) into 3D CAD software Fusion 360. This plugin generates 2D image inspiration for conceptual CAD and product design workflows. \nickname{} helps users craft text prompts by providing 3D keywords, design/styles, and parts from GPT-3. Users can also generate from image prompts based on a render of their current workspace, letting users use their 3D modeling progress as a basis for text-to-image generations.}
  \Description{A 3D model of a transparent car with a text-to-image AI generation is on the left side, which shows the workspace of Fusion 360, a 3D modeling software. The right side is the plugin \nickname, which has shows a palette of colored suggestions. These suggestions are different designs, styles, and parts of a car. There is a final prompt at the bottom of the plugin that says "3d illustration of a Batmobile mesh that has red details in the tires".}
  \label{fig:teaser}
\end{teaserfigure}

\maketitle

\section{Introduction}

%TC:ignore
% what is the problem statement? the problem is that designing 3d models is challenge, where do people get inspiration image search engines, video tuorials, 3d models. plus have to understand 
% how do you get inspiration for 3d modeling
%TC:endignore

Designing 3D models in CAD software is challenging---designers have to satisfy a number of objectives that can range from functional and aesthetic goals to feasibility constraints. Coming up with ideas takes a lot of exploration, even for experienced designers, so they often consult external resources for inspiration on how to define their geometry. They browse 3D model repositories~\cite{funkhowser_model_by_ex}, video tutorials, and image search engines to understand conventional designs and different aesthetics~\cite{frameworkforcad}. This process of conceptualizing CAD designs is pivotal to the product design process, yet few computational methods support it~\cite{collabcad, frameworkforcad}.

A recent innovation that can more directly provide inspiration to designers is text-to-image AI. Tools such as DALL-E \cite{dalle}, Imagen \cite{imagen}, Parti \cite{parti}, and Stable Diffusion \cite{Rombach_2022_CVPR} are AI tools that have the generative capacity to access and combine many visual concepts into novel images. Given text prompts as input, these tools can capture a wide variety of subjects and styles \cite{liu2021design}. In online communities, users have already developed methods to elicit images with 3D qualities \cite{promptbook, liu2021design} 
 by including prompt keywords such as ``3D render'' or ``CGI''. Recent advancements have also allowed users to interact with text-to-image AI systems by passing in image prompts, where images are used as prompts in addition to text. Generations can now be varied or built off of previous generations. These innovative functions make the integration of text-to-image AI within existing creative authoring software more feasible. 

However, how AI-provided image inspiration can contribute to CAD and product design workflows has not yet been fully explored. In this paper, we seek to understand how text-to-image AI can assist 3D designers with conceptual CAD and design inspiration and where in creative workflows designers can most benefit from AI assistance. Furthermore, we investigate how text-to-image tools respond to image prompts sent from 3D designers as they build up complexity in their designs. To do so, we integrated three large AI models---DALL-E, GPT-3, and CLIP---within Fusion 360, an industry standard software for computer-aided design (CAD). We implemented a plugin within the software which we call \textit{\nickname}. This plugin helps translate a designer's goals into multimodal (text and image) prompts which can produce image inspiration for them. After a designer inputs their goals (i.e. to design a "truck"), the plugin provides a number of related parts, styles, and designs that help users craft text prompts. These suggestions are drawn from the world knowledge of GPT-3 \cite{gpt3} to help users familiarize themselves with relevant design language and 3D keywords that can better specify the text prompt. The plugin interactively updates an image preview from the software viewport that shows an image prompt which can be passed into DALL-E~\cite{unclip}, giving users a direct bridge between their 3D design workspace and an AI model that can generate image inspiration. Additionally, having a lens on what the designer is actively working on allows the plugin to highlight what prompt suggestions may work best, which is implemented in the system by using CLIP \cite{radford2021learning} to approximate model knowledge. To evaluate \nickname and how well it can integrate into 3D workflows, we conducted a user study with thirteen users of Fusion 360 who spanned a variety of backgrounds from industrial design to robotics. \revAdd{ We found that \nickname can benefit CAD designers as a system that supports conceptual CAD, helps prevent design fixation, produces reference images, and inspires design considerations. }

We present the following contributions:

\begin{itemize}
    \item \nickname, a plugin that generates AI-provided image inspiration for CAD and product design by helping users craft text prompts with design language (different parts, styles, and designs for a 3D object) and image prompts connected to their work in progress.
    \item An exploratory user study (n=13) demonstrating text-to-image AI use cases in 3D design workflows and an analysis of prompting patterns and prompt complexity.

\end{itemize}
In our discussion, we propose prompt bibliographies, a concept of human-AI design history to track inspiration from text-to-image AI. We conclude on how text-to-image AI can integrate with existing design workflows and what can be best practices for generative design going forward.

\section{Related Work}

\subsection{Prompting}

Prompting is a novel form of interaction that has come about as a consequence of large language models (LLMs) \cite{gpt3}. Prompts allow users to engage with AI using natural language. For example, a user can prompt an AI, ``What are different parts of a car?'' and receive a response such as the following, ``Wheels, tires, and headlights''. These prompts give LLMs context for what tasks they need to perform and help end users adapt the general pretraining of LLMs without further finetuning \cite{gwern_2020, reynolds2021prompt}. By varying prompts, users can query LLMs for world knowledge, generative completions, summaries, translations, and so forth \cite{gpt3, opal}. Datasets around prompting are also beginning to emerge to benchmark generative AI abilities. PARTI \cite{parti} provides a schema and a set of prompts to investigate the visual language abilities of AI. Coauthor \cite{Lee_2022} provides a dataset of rich interactions between GPT-3 and writers. Audits of models have also been performed by collecting generated outputs of AI models at scale and conducting annotation studies, as in \cite{liu2021design} and \cite{initialimages}. As generative AI communities have gained momentum online, crowd-sourced efforts on Twitter and Discord have also organized to disseminate prompting guidance \cite{promptbook} that suggest experimentation with various style and medium keywords (e.g. ``isometric'', ``3D render'', ``sculpture'' etc.).

Recent research directions have begun to develop workflows around prompts. AI Chains \cite{aichains} studied how complex tasks can be decomposed into smaller, prompt-addressable tasks. Promptchainer \cite{promptchainer} unveiled an editor that helps users visually program chains of prompts. Prompt-based workflows were explored in \cite{promptbasedellen} to make prototyping ML more accessible for industry practitioners. Other systems have tested pipelines that concatenate LLMs with text-to-image models. In Opal \cite{opal}, a pipeline of GPT-3 initiated prompt suggestions generated galleries of text-to-image generations to help news illustrators explore design options in a structured manner. Similarly, a visual concept blending system in \cite{ge2021visual} used BERT \cite{devlin2019bert} to surface shape analogies and prompt text-to-image AI for visual metaphors. A key finding from Opal and the visual blends system \cite{ge2021visual} that we apply in \nickname is that LLMs can help generate prompts so end users can efficiently explore design outcomes.

New modes of prompting have also started to emerge. Users can now pass in image prompts and have AI models autocomplete images and canvases in methods called inpainting and outpainting \cite{dalle,outpainting}. These functions have been implemented within state-of-the-art text-to-image AI systems \cite{dalle}. \nickname{} is the first to systematically generate image prompts from CAD software (Fusion 360) and help users incorporate their 3D design progress into text-to-image generations.

\subsection{Generative Models}
Generative AI models have long been excellent at image synthesis. However, many early models were class-conditional, meaning that they were only robust at generating images from the classes they were trained on \cite{karras2020analyzing, stylegan, Attngan, mirrorgan, leicagan, tedigan}. The most recent wave of generative AI models can now produce images from tens of thousands of visual concepts due to extensive pretraining. CLIP \cite{radford2021learning}, a state-of-the-art multimodal embedding, was trained off of hundreds of millions of text and image pairs, giving it a broad understanding of both domains. The pretraining of CLIP has also helped it serve as an integral part of multiple generative workflows \cite{murdock2021big, crowson,vqgan, crowson2022vqgan}  and training regimes \cite{clipforge, clipcap}. Large open-source efforts had previously paired CLIP with GAN models, using it as a discriminator to optimize generated images toward text prompts. 
%TC:ignore
\revDel{Before the popularity of text-to-image tools, trends in AI-generated art included the computational exploration of latent spaces \cite{ghosh2019interactive, ganslider}.}
%TC:endignore
 The novelty of generating media through language has brought many text-to-image tools into production such as Midjourney, DALL-E, and Stable Diffusion. DALL-E \cite{unclip} demonstrated how CLIP embeddings can help generate images with autoregressive and diffusion-based approaches. Diffusion is key within many of the aforementioned methods to increase the quality of text-to-image outputs \cite{murdock2,crowsondiffusion, dalle}. New text-to-image approaches have led to more diverse methods of user interaction. Make-a-Scene \cite{meta_ai} allows users to interact with generations by manipulating segmentation maps, and DALL-E gives users the ability to paint outside the edges of an image, allowing for unlimited canvases \cite{outpainting}. Textual inversion~\cite{Rombach_2022_CVPR} gives users the ability to train and trade novel concepts learned by the AI \cite{textualinversion} off of a few examples. These models have extraordinary generative capacity, but their ability to be used nefariously has also inspired new approaches to safeguarding AI outputs from redteaming~\cite{redteaming} to large scale audits for social and gender biases~\cite{dall-eval}. %TC:ignore
 \revDel{With \nickname, we present methods to narrow the scope of general-purpose generative models (GPT3, DALL-E), such that they are useful in a specific domain: 3D design.}
 %TC:endignore

Text-to-3D methods such as CLIP-Sculptor, DreamFusion, and Point-E \cite{dreamfields, dreamfusion, textcraft, clipforge} also exist and are rapidly improving, but they have far longer inference times \cite{dreamfields} and required computing power \cite{dreamfields}. They are also often constrained to producing shapes that are limited in diversity \cite{textcraft}, fidelity \cite{textcraft, clipforge}, stylistic range \cite{pointe}, and capabilities for variable binding owing to the smaller volume of paired text-shape data online \cite{ textcraft}.
Advances using diffusion models as a prior have also made the generation of complex, textured 3D models possible \cite{dreamfusion}. However, text-to-3D approaches result in scene \cite{dreamfusion}, voxel \cite{textcraft,clipforge}, pointcloud \cite{pointe}, and mesh \cite{dreamfusion} representations that are medium or high fidelity from the get-go. This can start a designer off at an unfamiliar stage in their workflow (with a medium or high fidelity geometry they might not know how to edit) or with a representation they do not usually use for CAD. To support conceptual CAD from the earliest stages possible, we investigate text-to-image rather than text-to-3D in \nickname{} as the most suitable starting point for AI-provided inspiration. We elaborate on how designers often start in 2D and build up to 3D forms using shape operations in Section 2.4.

\subsection{Creativity Support Tools}

Human-computer interaction research on creativity support tools has long showcased ways to facilitate text-based content creation. Early systems showed that users could iteratively define images based on chat and dialogue \cite{chatpainter, keepdrawingit}.  AttriBit \cite{Chaudhuri:2013:ACC} allowed users to assemble 3D models out of parts matched on affective adjectives. Sceneseer \cite{sceneseer} and Wordseye \cite{wordseye} allowed users to create scenes via sentences. However, since the advancement of AI tools, much of the momentum has now concentrated around human-AI co-piloted experiences. Systems such as Opal \cite{opal}, Sparks \cite{sparks}, FashionQ \cite{fashionq}, and the editors in \cite{stolenelephant} are examples of AI-assisted ideation. In tandem, many frameworks for computational creativity \cite{llano} and human-AI interaction \cite{humanaiguidelines} have cropped up to understand concerns such as ownership and agency when AI is involved in the creative process. Gero et al. \cite{gero2020} found that users can establish better mental models of what AI can and cannot do if they have a sense of its internal distribution of knowledge.

Practices for creativity support tools that we revisit from an AI perspective include the idea of design galleries \cite{designgalleries}, timelines and design history \cite{chronicle}, natural language exploration \cite{datatone}, and collaboration support \cite{socialglue}. DataTone \cite{datatone} demonstrated how interactive prompting with widgets can help build specificity in a text-based interface. Suh et al. \cite{socialglue} demonstrated that AI-generated content could facilitate teamwork within groups by helping establish common ground between collaborators. 
While many systems have been built with generative AI capabilities \cite{cococo, cobuild, ghosh2019interactive} and even for text-to-image workflows \cite{opal}---none that we know of have applied text-to-image AI for 3D design workflows.

\subsection{CAD Conceptual Design and Workflows}

%TC:ignore
\revDel{
For example, to create a mug or a table (consumer goods commonly designed in CAD), they might extrude a circular cross section along an arc to make the handle of a mug or loft four squares upwards to create the legs of a table. 
}
%TC:endignore

%conceptual cad
CAD is a highly complex design activity that usually involves a significant amount of conceptual design, as later stages of prototyping can incur material costs. Because CAD evolved in part from 2D drafting, CAD often relies on 2D representations such as freehand drawing and computer-assisted sketches \cite{Hartman2004DefiningEI, dreamsketch, gesturespeechcad}. In these early stages, designers are also gathering inspiration from external sources like 3D model repositories \cite{funkhowser_model_by_ex} (e.g. Onshape, Google Poly), video tutorials, and reference images \cite{photo-inspired_2011} to inform their sketches. Users operate over these 2D representations (sketch profiles and planes) to apply constraints and dimensions and to take their models into 3D using operations such as extrusion, lofting, revolving and so on \cite{Hartman2004DefiningEI}. It has been found that early stage CAD and product design ``tends to be ambiguous, incomplete, and expressive with high levels of uncertainties''~\cite{dreamsketch}, and there is less focus on constraints and parameters~\cite{gesturespeechcad, cadoncreativity}. Conceptual CAD also can involve text and image exploration; mechanical engineers perform system decomposition to understand model needs, and industrial designers collect moodboards and perform market research~\cite{collabcad}.

%TC:ignore
\revDel{
Conceptual CAD presents in many different disciplines of CAD from industrial design to mechanical engineering as a process that is fundamental to support not only within each discipline but across disciplines \cite{collabcad}. 

Conceptual CAD can vary depending on the discipline and operate under different frameworks: system decomposition (mechanical engineers), moodboards / market research (industrial designers), analytic diagrams (architects) \cite{collabcad}.

Designers can also start by remixing existing parts \cite{datadriven_3d,funkhowser_model_by_ex,attribit} found using tools like Thingiverse or Sketchfab.

Generating many ideas is important because it has also been found that going to CAD too early in the process can lead to design fixation, which is when designers become more resistant to changes as they invest time into their design \cite{cadoncreativity,designfixation}.}
%TC:endignore

%workflows
\revAdd{ One direction within HCI work has focused on capturing and understanding CAD workflows. Screencast \cite{screencast,chronicle} collects timelines of authoring operations from CAD help forums. From Screencast data, workflow graphs \cite{chang2020workflow} have been proposed as a way to characterize 3D modeling workflows. These graphs have shown that users can arrive at 3D models through different paths. For example, to design a mug, a user can design in parts and in interchangeable sequences; they can first create the body of the cup, and then the handle, or vice versa. Examinations of CAD experts have also generalized CAD modeling as procedures of increasing detail, working from sketches to geometric forms to finishing features \cite{Hartman2004DefiningEI}. }

%closest work
\revAdd{Prior work on applying generative models and AI for knowledge-based design in CAD and industrial engineering does exist \cite{krahe21aibased, mirra22, Gero1993ModelingCA,liao2020ai}. Liao et. al. note that parametric CAD tools do not offer ``cognitive supports for search nor highlight new information a designer might not have thought of'', which is where generative AI can assist by providing triggers for novel solutions \cite{bernal2015}. The closest works to \nickname{} would be DreamSketch \cite{dreamsketch} and Dream Lens \cite{dreamlens}, systems for generative design exploration. DreamSketch, helped explore 3D design ideas by passing in sketches, design variables, and constraints that retrieved generative designs from topology optimizers. Dream Lens helped users explore and visualize large-scale generative design datasets based on parameters. Rather than freehand sketches or parameters, \nickname{} presents a method for supporting conceptual CAD through text-based exploration of design knowledge and text-to-image generations. 
}

%TC:ignore
\revDel{We deploy 3DALL-E to provide a novel system for conceptual CAD, field test text-to-image technology with CAD designers, and understand how generations can influence 3D workflows and inspire product designs.}
%TC:endignore

\section{Designing with \nickname}

\subsection{Design Rationale}

\revAdd{Engaging with text-to-image AI means coming up with many prompts. Users have to exhaustively experiment with AI to see what words it can understand and render well. To streamline prompt ideation for a CAD environment, \nickname{} helps users efficiently assemble 3D design knowledge into prompts. For example, for a table, a user may know common designs like \textit{``dining table''} or \textit{``desk''} but may otherwise not know design vernacular (\textit{``lift-top'', ``drop-leaf'',} or \textit{``nesting''} table) that \nickname can efficiently supply.}

\subsection{The \nickname interface}

\begin{figure*}[h]
    \centering
    \includegraphics[width=0.90\textwidth]{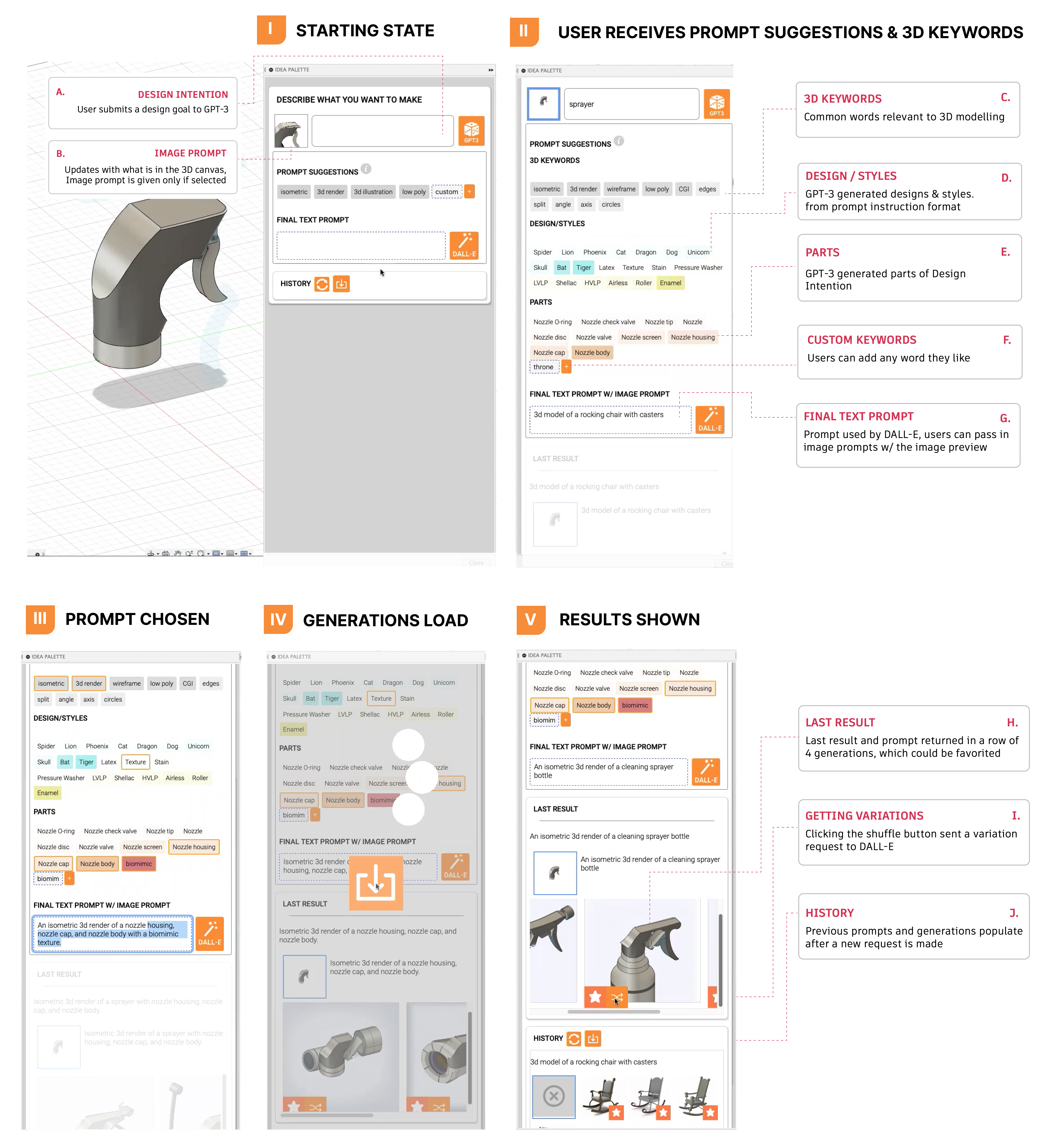}
    \caption{\nickname walkthrough. Step I: Initial state, where users can type their design intentions. Step II: Users are presented with prompt suggestions from GPT-3. Step III: Selected suggestions are rephrased into an editable prompt. Step IV: Users wait as DALL-E generates. Step V: Results are shown. A cursor hovers over a shuffle icon, which is how users can launch variation requests from DALL-E. }
    \Description{ Comic strip layout of user interface screenshots going left to right showing stages users experience as they use 3DALL-E. The first shows an empty starting state, where none of the fields have populated. The second shows a palette of colored pill options. The third shows a similar palette but with prompt filled in a final text box. The fourth shows a loading page. The fifth shows the plugin returning results from the text-to-image AI. Stage 1: a default state, where users can type their design intentions. Stage 2: Users are presented with prompt suggestions from GPT-3. Stage 3: selected suggestions are rephrased into an editable prompt. Stage 4) Users wait as DALL-E generates. Stage 5) Results are shown. A cursor hovers over a shuffle icon, which is how users launched variation requests from DALL-E.}
    \label{fig:timeline}
\end{figure*}

\nickname is provided as a panel on the right hand side of the 3D workspace (Fig.~\ref{fig:teaser}). Fig.~\ref{fig:timeline} shows the steps users go through when designing with \nickname inside their 3D workspace and presents the main interface components. \nickname allows users to construct prompts relevant to their current 3D design, which can then be sent to DALL-E to retrieve AI-provided image inspiration. Once generations are received, users are able to download them, see a history of previous results, and create variations of generations that they want to explore more from. In what follows, we will present these different steps with a short walkthrough.

\subsection{Constructing Text Prompts for AI-Provided Inspiration}

Users begin at the starting state shown in Fig.~\ref{fig:timeline}-I, where they can describe what they want to make by typing in their goal (Fig.~\ref{fig:timeline}A). Once they do that, different prompt suggestions populate the sections with 3D keywords, designs/styles, and parts (Fig.~\ref{fig:timeline}-II). These suggestions help steer the generations toward results relevant to 3D modeling as well as provide design language a user might otherwise not be familiar with. For example, querying a chair could return a series of existing designs such as an egg chair, an Eames chair, or a Muskoka chair, helping familiarize the user with the design language befitting of chairs. Once users select a set of prompt suggestions (e.g. \textit{``3d render, isometric, plant stool, wrought-iron''}), an automatically rephrased prompt appears in the final prompt box (e.g. \textit{``isometric 3d render of a wrought-iron plant stool``}) as shown in Fig.~\ref{fig:timeline}-III. This prompt is still editable by the user, and a text box to add custom keywords is also available when clicking the orange `+' button in the parts section (Fig.~\ref{fig:timeline}F).

Prompt suggestions (Fig.~\ref{fig:timeline}C--E) are color-coded with a color for the group they belong to (blue for designs, green for styles, orange for parts) and varied in opacity to indicate how strongly their text aligns with the image prompt (see Fig.~\ref{fig:texthighlights} for implementation details). For example, from a set of styles like ``mid-century modern, contemporary, and art deco'', if ``art deco'' was most strongly highlighted (i.e. more opaque --  darker green), it meant that the image prompt had the greatest probability of being matched with ``art deco''. \revAdd{\nickname{} suggests keywords to elicit 3D qualities particular to 3D models and renders, following design guidelines from related work \cite{promptbook, liu2021design}. Styles are suggested to allow users to steer the aesthetic language of their generation and engage with inspiration spanning different time periods, traditions, and mediums (e.g. \textit{``mid-century modern'', ``Brutalist''}, or \textit{``CGI''}). Using style keywords is also a recommended tip from prior work and existing AI systems \cite{liu2021design, dalle, promptbook}. \nickname{} suggests parts as 3D models are often assemblies of parts, as established in work on part-based authoring systems~\cite{attribit} and part datasets~\cite{abc, partnet,willis2020fusion}.  Other dimensions like material and function could have been explored without loss of generality. However, we chose to focus on geometry-relevant suggestions instead of appearance (material) or abstract goals (function).}

\subsection{Crafting an Image Prompt}
Users can also choose to include an image that is automatically extracted from their current 3D modeling workspace in addition to their text prompt (image+text prompt) or choose to exclude it (text-only prompt). Image prompts are only passed in when users select the image preview (Fig.~\ref{fig:timeline}B), making it active. Using the 3D software to render the viewport allows \nickname to programmatically deliver clean prompts without tasking the user with any erasing or masking. Users can easily toggle the visibility of certain parts of their model using the 3D software's built-in functionality and request for \mbox{DALL-E} to fill in the details for those hidden parts. %An example of an image prompt is seen in Fig.~\ref{fig:system_design}.

%\subsection{Functionalities from DALL-E}
\subsection{Receiving DALL-E Results and Retrieving Variations}
Once the user is satisfied with the prompt, they click the DALL-E button next to the final text prompt (Fig.~\ref{fig:timeline}G) to generate either a text-only or image+text prompt (depending on whether the image preview is selected). While waiting for results (Fig.~\ref{fig:timeline}-IV), the user is shown a spinner animation. When the results are ready, the user can click the orange download button to pull the results from DALL-E into the \nickname{} interface. 

Results are returned in sets of four (Fig.~\ref{fig:timeline}-V). When the user hovers over a result, they are presented with a menu that allows them to `star' their favorite results and click the `shuffle' button to get more \textit{variations} on that particular result (Fig.~\ref{fig:timeline}I). 
\revAdd{ These are retrieved using DALL-E's built-in functions that generate similar images given an image input.} Lastly, \nickname also keeps a history of previous generations (Fig.~\ref{fig:timeline}J). 

\section{System Implementation}

\begin{figure*}
    \centering
    \includegraphics[width=\textwidth]{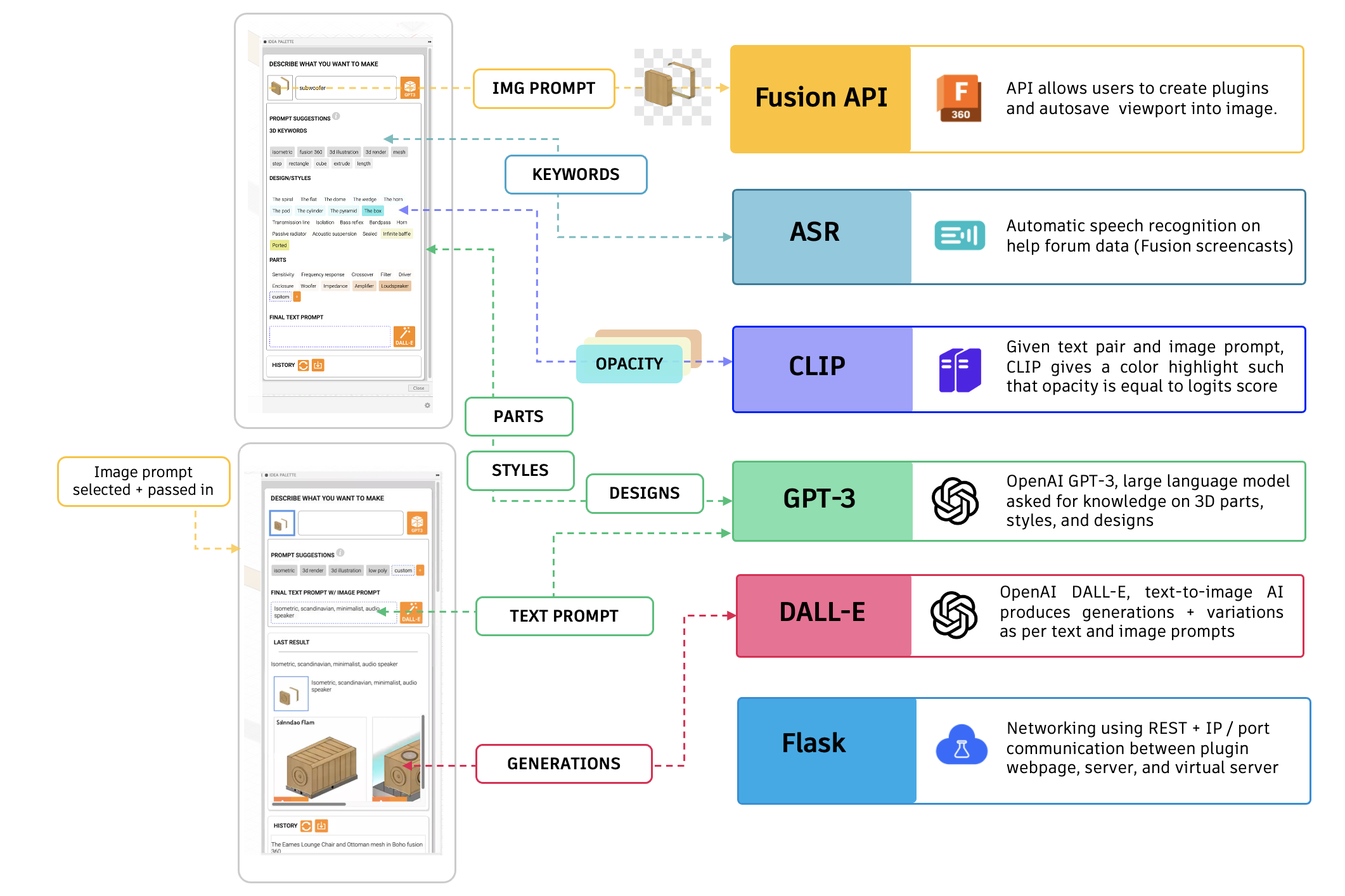}
    \caption{System design showing the architectures involved in \nickname, which incorporates three large AI models into the workbench of an industry standard CAD software. In the top left panel, we show how text AI outputs are displayed in the UI. In the bottom left panel, we show how users could pass in image prompts and retrieve DALL-E generations within the plugin. }
    \Description{ System design figure where each colored bar represents a different element of \nickname, which incorporates three different large AI models into the workbench of an industry standard CAD software. In the top left panel, we elaborate the architecture and networking communication that provides prompt suggestions. In the bottom left panel, we illustrate how users could pass in image prompts and retrieve results from DALL-E within the plugin. }
    \label{fig:system_design}
\end{figure*}

\nickname was implemented within Autodesk Fusion 360 \cite{fusion_360_api} as a plugin and written with the Fusion 360 API, Python, Javascript, Selenium, and Flask. Fig.~\ref{fig:system_design} illustrates how we embedded DALL-E, GPT-3, and CLIP into one user interface. All actions in \nickname{} were logged by the server to facilitate analysis of participant behavior in the study (Sect.~\ref{sec:evaluation}). Note that \nickname could be implemented generically in most 3D modeling tools. The needed functionality from Fusion 360 is relatively basic: a custom plugin system and ways to render the viewport as an image.

Prompt suggestions were populated by querying the GPT-3 API for the following: \textit{``List 10 popular 3D designs for \{QUERY\}? 1.'', ``What are 10 popular styles of a \{QUERY\}? 1.'', and ``What are 10 different parts of a \{QUERY\}? 1.''.} \revAdd{ These queries were split using regular expressions such that each suggestion was one button on the interface. } To rephrase chosen suggestions, GPT-3 was prompted: \textit{``Put the following together: \{SUGGESTIONS\}''}.

Ten 3D keywords are sampled from a set of high frequency words (n=121) in a Fusion 360 Screencast dataset. Screencasts are videos used to communicate help and tutorials in forums \cite{screencast,chronicle}. Automatic speech recognition (ASR) of these videos produced transcripts; these transcripts were processed with \revAdd{standard count vectorization using NLP modules from Sklearn}, filtered out for general purpose words \revAdd{ (words that were not specific to CAD)}, and sorted by frequency to get the final keywords set.

\begin{figure*}
    \centering
    \includegraphics[width=\textwidth]{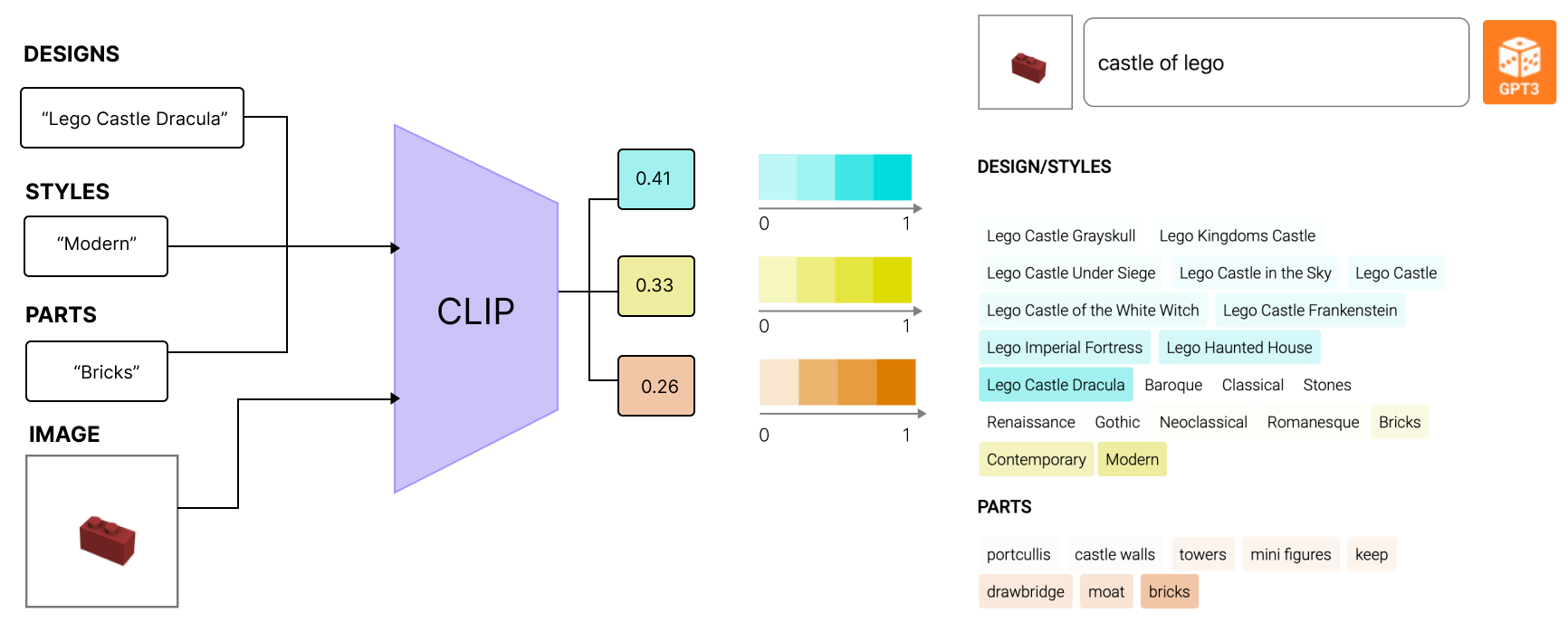}
    \caption{Diagram showing how text highlights were calculated using CLIP with image and text from the prompt suggestions as input. The CLIP logits score was set as the opacity of each prompt suggestion. Each type of suggestion was colored differently.}
    \Description{ 3D modeling workspace shown on left with lego in view. A legend at the top shows how prompt suggestions were colored blue, green, or orange from CLIP. The plugin is on the right showing an example of prompt suggestions returned given an image prompt of a lego and a text query, "castle of legos". }
    \label{fig:texthighlights}
\end{figure*}

Text highlights were calculated by passing each of the prompt suggestions and the image prompt to CLIP, which was hosted on a remote server. CLIP produces \revAdd{softmaxed }logit scores\footnote{\revAdd{Applying softmax to logit scores yields normalized linear probabilities.}} that suggest how similar each text option was to the image, a value \nickname renders as the opacity of each highlight.  The stronger the highlight, the greater the probability a text option matched what a user had in their viewport. DALL-E was trained with CLIP text and image embeddings. By using CLIP's embedding in this way, users receive a computational guess for how well DALL-E might be able to interpret each prompt suggestion, while also dialing down the options they need to focus on (Fig.~\ref{fig:texthighlights}). The 3D keywords were by default gray, while designs, styles, and parts were matched to gradations of blue, green, and orange respectively.

We used the Fusion 360 API to automatically save the viewport to a PNG image every 0.3 seconds. The workspace of Fusion 360 (the gridded background pictured in Fig.~\ref{fig:teaser}) was rendered transparently in the PNG image.

\section{Evaluation}\label{sec:evaluation}
Implementing \nickname{} within Fusion 360 gave us a focused application context to evaluate text-to-image AI within a creative workflow.  We set out to investigate the following research questions for \nickname to understand in what ways text-to-image AI can be useful for 3D designers.
 \begin{itemize}

 \item \textit{Generation Patterns within Workflows.} Are there certain patterns to how CAD designers use text-to-image generations within their workflows, and do these patterns differ depending upon the 3D modeling task?

 \item \textit{Assisted Prompt Construction.} How helpful are different features (prompt suggestions, CLIP highlighting, automatically captured viewport images) for the construction of text and image prompts?

\item \textit{Prompt complexity.} How many concepts do people like to put within prompts?

\end{itemize}

To do so, we conducted an exploratory study with 3D CAD designers (n=13, 10 male, 3 female). Participants were recruited from internal channels within a 3D design software company as well as through a design institute mailing list at a local university. Participants were compensated with \$50 dollars for 1.5 hours of their time. The average age of the participants was 28, and they had an average of 4.13 years of experience with Fusion 360 (min=1 year, max=8 years). Five had experience with the generative design environment within Fusion, and three had prior experience with AI / generative art systems. The participants spanned a range of disciplines from machining to automotive design. Domains of expertise, frequency of use, and years of experience with the 3D software are listed in Table \ref{tab:participants}. Based on the system implementation in a CAD software, we focused on CAD designers and product designers rather than 3D artists and 3D concept art more broadly.

\subsection{Experimental Design}

Participants were given two different 3D modeling tasks: $T_{edit}$ to edit an existing model and $T_{create}$ to create a model from scratch. The intention of having these two tasks was to show how \nickname{} might affect creative workflows at different stages of the 3D modeling process. The ordering of these tasks was counterbalanced to mitigate learning effects. This experimental design was approved by a relevant ethics board.

Before the study, participants were sent an email with DALL-E's content policy to disclose that they were going to use AI generative tools. During the study, participants were given a brief introduction to the different AI architectures involved (GPT-3, DALL-E) and given two general tips on prompting: 1) text prompts should include visual language, 2) text prompts are not highly sensitive to word ordering \cite {liu2021design}. Participants were then given a walkthrough of the user interface and the different ways they could generate results from GPT-3 and DALL-E. The study was conducted virtually via Zoom and through remote control of the experimenter's Fusion 360 application and plugin.

\begin{figure*}
    \centering
    \includegraphics[width=\textwidth]{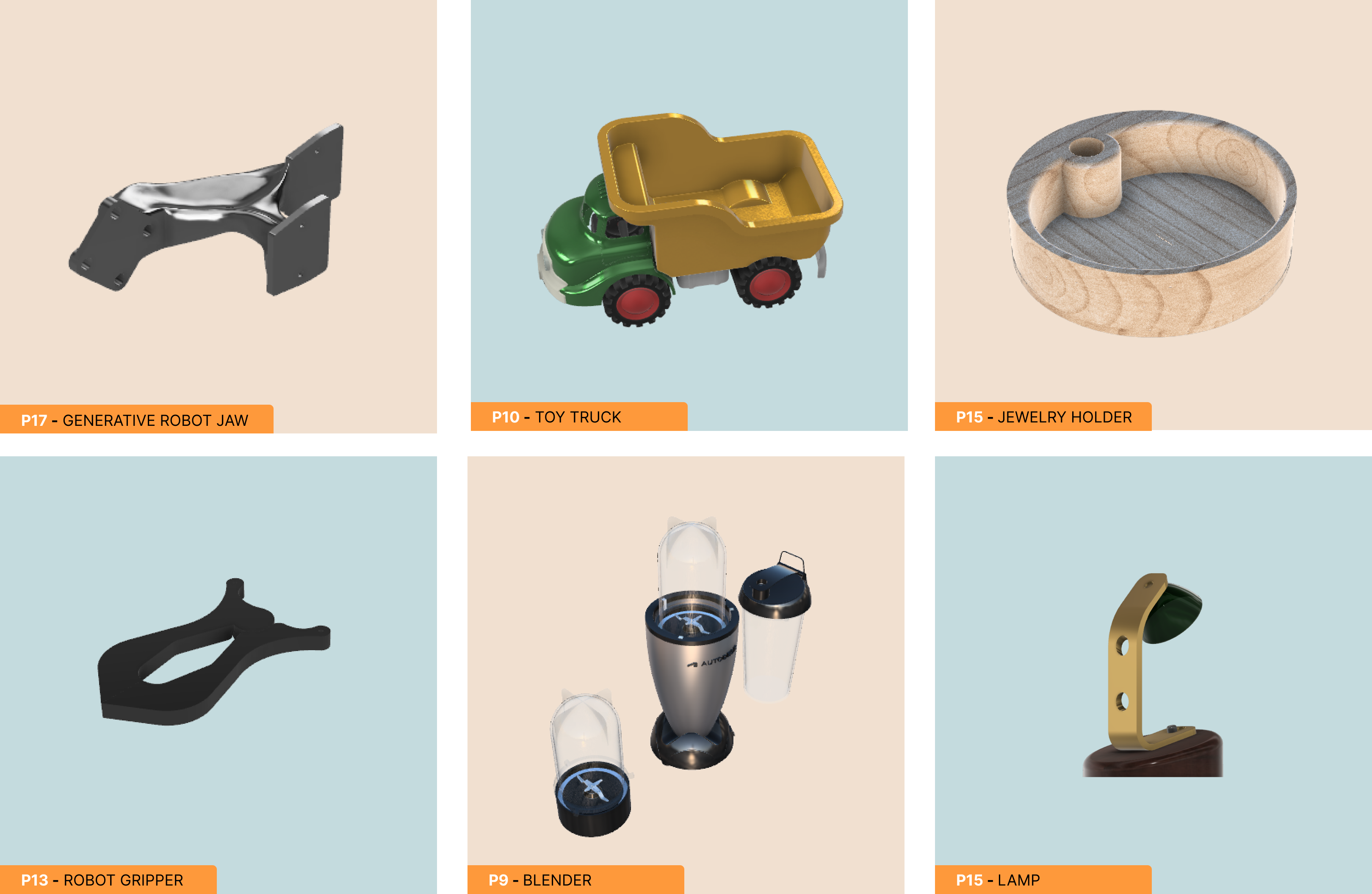}
    \caption{Examples of 3D designs participants brought in during $T_{edit}$, which was to edit an existing model.}
    \Description{2x3 grid of 3D designs participants brought in. These included a generative robot jaw, toy truck, jewelry holder, robot gripper, blender, and lamp.}
    \label{fig:examples_a}
    \vspace{0.6cm}
\end{figure*}

$T_{edit}$ was to modify an existing 3D model that the participants had brought with them to the study. Participants were told to bring a non-sensitive model, meaning one that did not include corporate data. There were no constraints on what the model could have been. Examples of models brought in can be seen in Fig.~\ref{fig:examples_a}. When a participant did not have a model to use, a random design was provided from the software's example library. This was the case for only one participant (P15).

\begin{figure*}
    \centering
    \includegraphics[width=\textwidth]{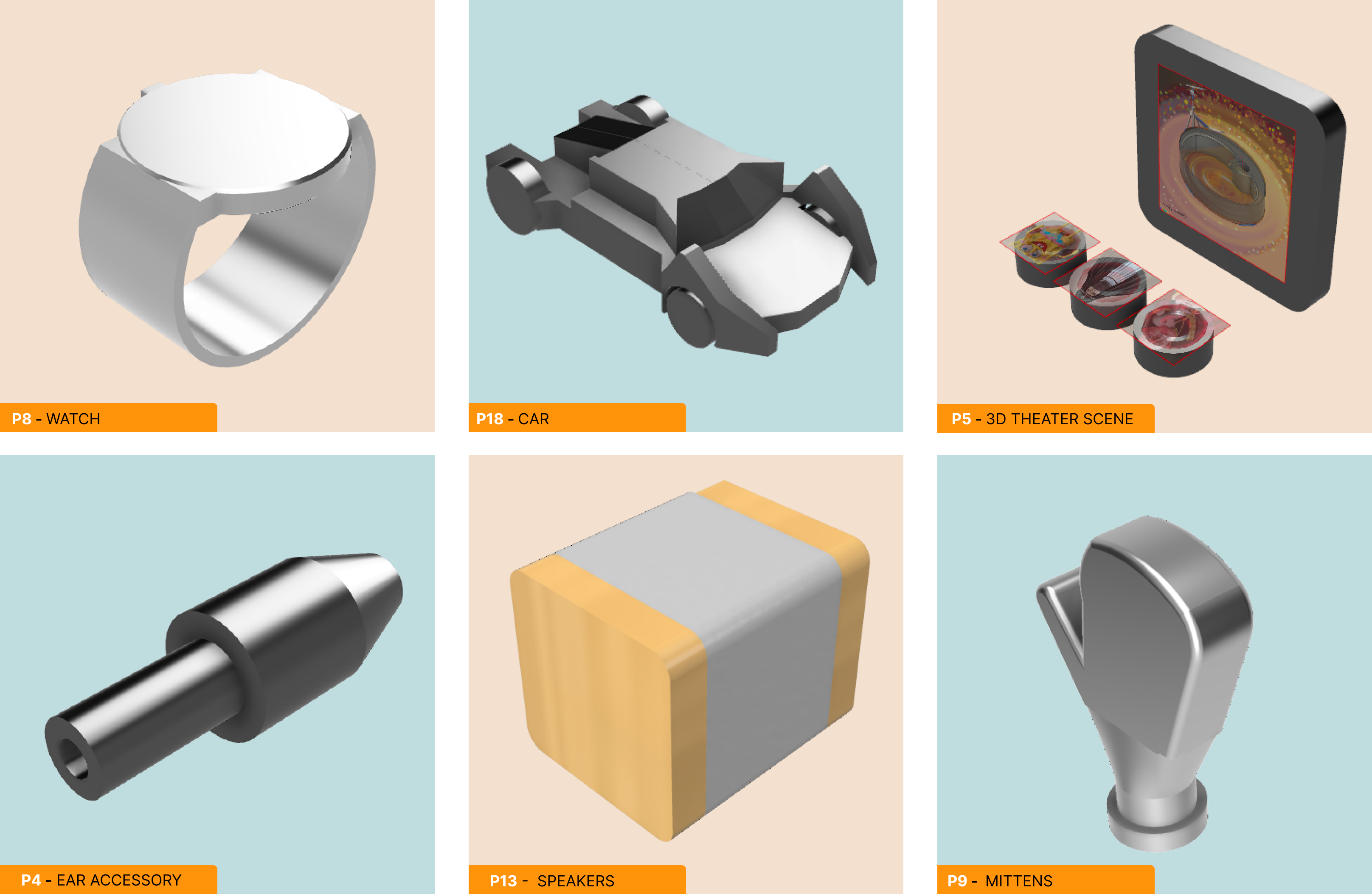}
    \caption{Example of 3D designs participants came up with during $T_{create}$, which was to create a model from scratch.}
    \Description{2x3 grid of 3D designs participants brought in. These included a watch, car, outdoor scene, ear accessory, speakers, mittens.}
    \label{fig:examples_b}
\end{figure*}

For $T_{create}$, participants were allowed to pick whatever they wanted to design from scratch. For each task, participants had 30 minutes to work on their model with the assistance of \nickname. \revAdd{We justify this duration of 30 minutes as a sufficient length of time based on prior work: DreamSketch \cite{dreamsketch} (30 to 60 minutes for 3D artifact creation) and Dream Lens \cite{dreamlens} (25 minutes for generative design exploration).} At the halfway point, participants were reminded of the time remaining and of any generation actions that they had not tried out yet from GPT-3 (prompt suggestions) or DALL-E (text-only prompts, image+text prompts, variations). Beyond this reminder, they were guided only if they needed assistance accomplishing something in the user interface. Examples of what participants created for $T_{create}$ can be seen in Fig.~\ref{fig:examples_b}. At the 30 minute mark, designers were told to wrap up their design.

\color{red}
\begin{figure}
    \centering
    \includegraphics[width=0.95\columnwidth]{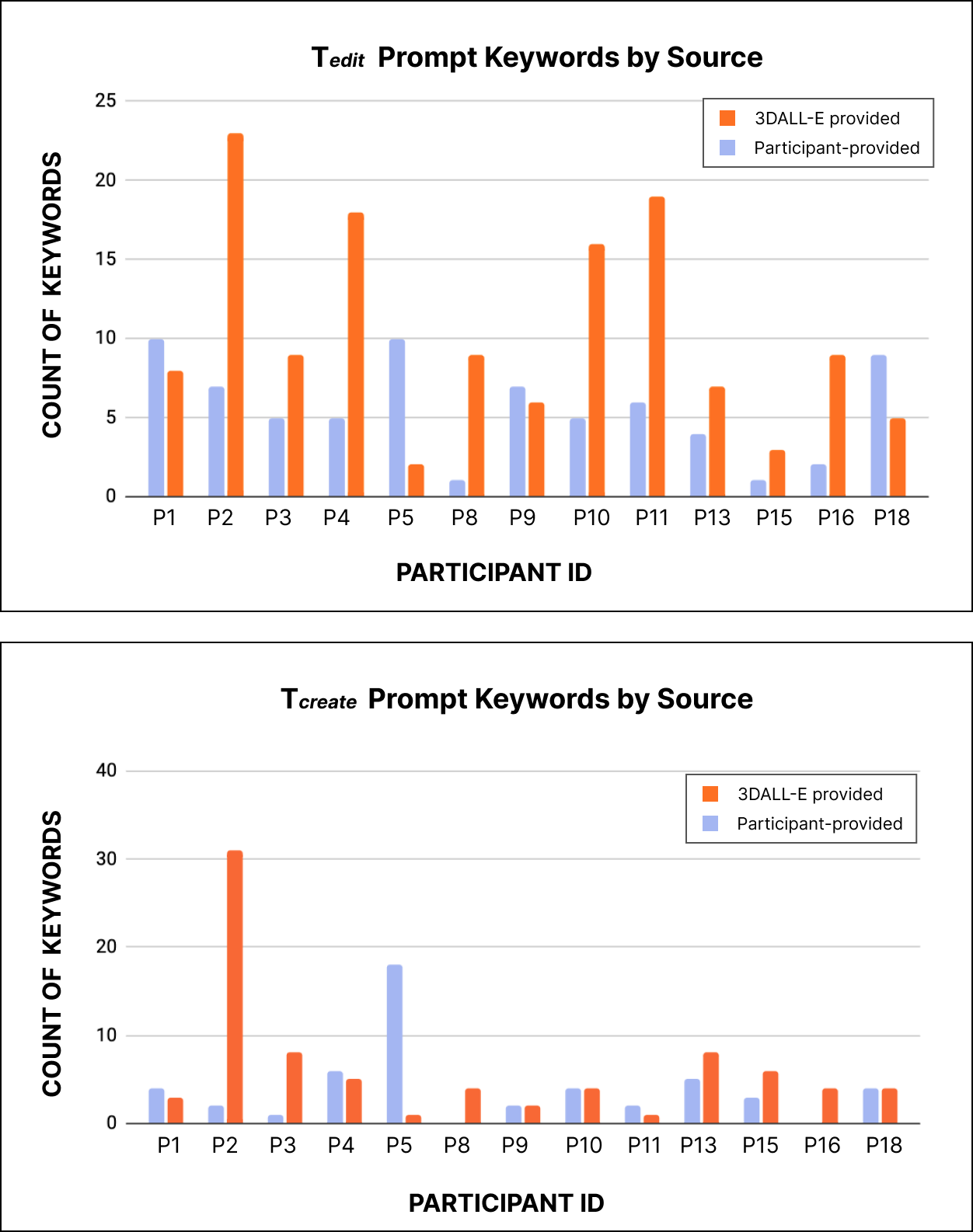}
    \caption{\revAdd{Count of prompt keywords by source (\nickname{}- or participant-provided) for each participant during $T_{edit}$ (top) and $T_{create}$ (bottom). \nickname{} provides at least half of prompt keywords for 9/13 participants in both tasks.} }
    \Description{Two graphs. Top graph has 13 sets of bars colored orange and purple. Orange bars represent count of prompt keywords provided by 3DALL-E, purple bars represent count prompt keywords provided by participants. The orange bars are higher across all participants, meaning the majority of prompt keywords come from 3DALL-E. }
    \label{fig:tcreate_promptkeywords}
\end{figure}
\color{black}

After completing each task, participants marked generations in their history that they felt were inspiring and completed a post-task questionnaire, which included NASA-TLX~\cite{nasatlx}, Creativity Support Index (CSI)~\cite{cococo,csi}, and workflow-specific questions. These questions can be found in the supplementary material. A semi-structured interview was conducted to understand their experience. 

\begin{table*}[!h]
  \caption{Table of participant details, with discipline, Fusion360 usage frequency, and years of experience. We list labels for the model they designed during $T_{create}$ and labels for the model they brought in ($T_{edit}$).}
  \Description{
Table of participant details, with discipline and Fusion 360 usage frequency, and years of experience. We list labels for the model they designed during T-create and labels for the model they brought in T-edit
  }
  \begin{tabular}{llllll}
    \toprule
    ID & Discipline    &  Fusion360 Freq\revAdd{.}  &  \revAdd{Exp.} & $T_{edit}$ & $T_{create}$ \\
    \midrule
P1   &   Mech. engineering,\revAdd{ CAD for robotics competitions } &    Few times /week &   \revAdd{4 yrs} &      robot &    prosthetic hand                   \\
P2   &    Design \revAdd{grad student, CAD + drone design instructor } &    Daily  &   \revAdd{4 yrs} &  drone    &             airplane \\
P3   &   \revAdd{ Mech. engineering + }design \revAdd {student, CAD hobbyist }  &   Few times /year & \revAdd{1 yr} &                 ring &                    iPhone\\
P4   &     Technical CAD \revAdd{software demos} and sales &   Daily&  \revAdd{7 yrs} & machined part &             Bluetooth     ear gauge \\
P5  &   Mech. engineering \revAdd{student, CAD hobbyist }  &   Few times /month &  \revAdd{2 yrs} &  jewelry holder base &              outdoor 3D scene\\

P8   &      Mechanical engineer  &             Few times /month &             \revAdd{2.5 yrs} &      \revAdd{table}  top &               table \\
P9  &        Mech. engineering student,  \revAdd{CAD hobbyist } &  Few times /year & \revAdd{2 yrs} &     spray bottle \revDel{mechanism} &         mittens\\
P10   &     Technical accounts executive \revAdd{for CAD (demos) } &    Few times /week & \revAdd{8 yrs} &             truck  &                     shelf \\
P11   &       \revAdd{ CAD }technical support for machining &              Daily & \revAdd{1.3 yrs} &               blender &                  bottle\\

P13   &  \revAdd{ CAD }software engineer, \revAdd{prev.} industrial designer  &     Daily &  \revAdd{8 yrs} &               gripper  &     speakers      \\
P15   &  Technical product manager  \revAdd{at car company }&    Few times /year & \revAdd{5 yrs}&                lamp &          bookshelf \\
P16 & Mechanical engineer & Few times /year & \revAdd{1 yr} &sensor mount & screwdriver \\
% P17   &      Shop supervisor &     Med / Med &                  threat multiplier &          environmental health\\
P18   &    Technical sales \revAdd{(CAD demos)}, industrial designer  &     Daily  & \revAdd{8 yrs} &                  microphone stand &          car\\
  \bottomrule
\end{tabular}
  \vspace{0.6cm}
  \label{tab:participants}
\end{table*}

\begin{figure*}
    \centering
    \includegraphics[width=0.70\textwidth]{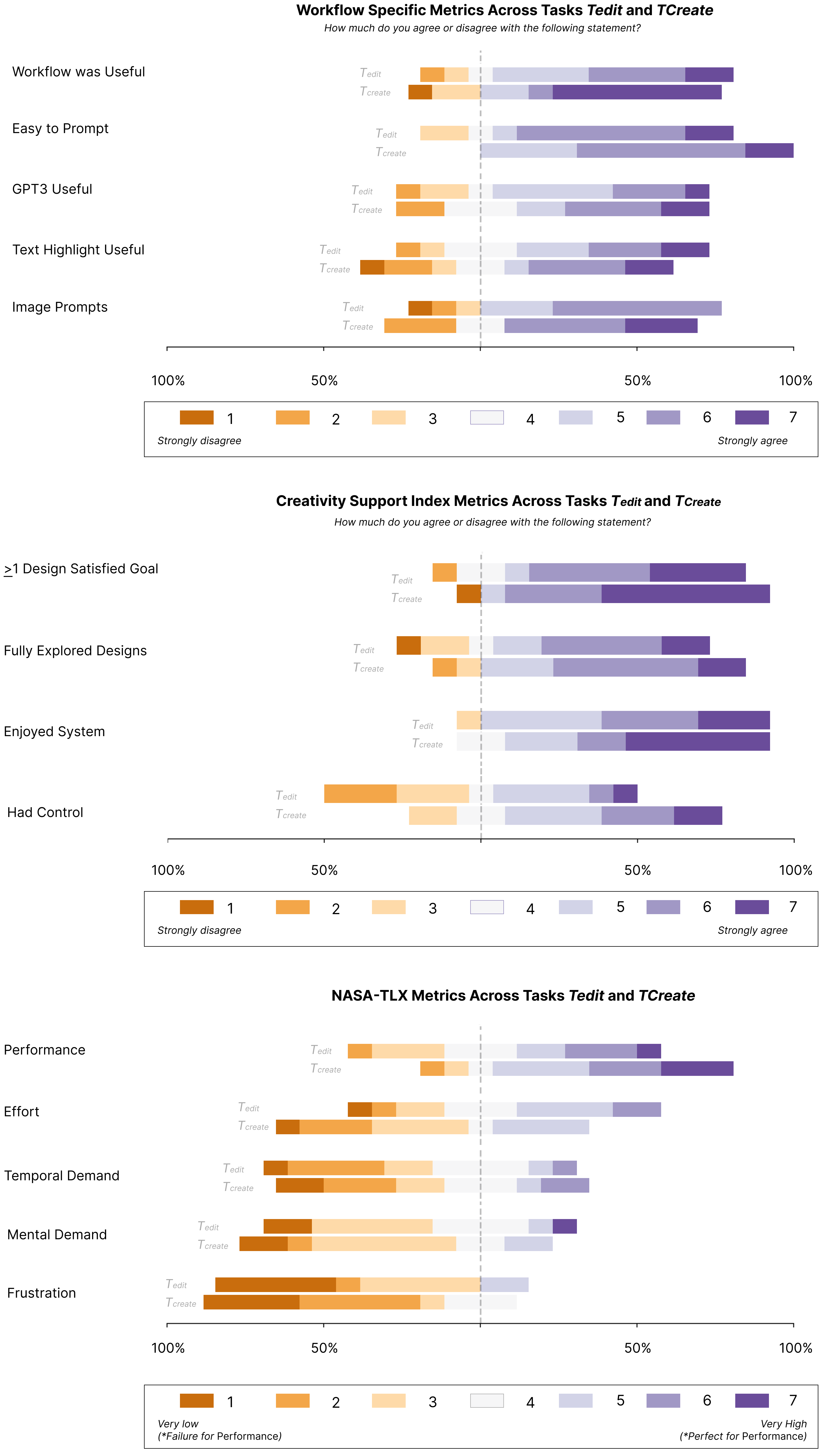}
    \caption{Distribution of Likert scale responses on NASA-TLX, creativity support index, and workflow-specific questions across all participants for both $T_{edit}$ and $T_{create}$. Full questions are in the Appendix. }
    \Description{
    Thirteen diverging bar charts split into three categories (NASA, CSI, and specific questions). Each bar chart has two bars, the top corresponding to the Task T-edit, and the second to T-scratch. Descriptions and interpretations can be found in the evaluation under Feedback. 
    }
    \label{fig:quant_metrics}
\end{figure*}

% \begin{figure}
%     \centering
%     \includegraphics[width=0.45\textwidth]{fig/fig_specific.png}
%     \caption{Distribution of Likert scale responses on specific questions relating to \nickname's workflow. There is a high concentration of responses in the High and Very High categories.}
%     \label{fig:my_label}
% \end{figure}

\subsection{\revAdd{Quantitative} Feedback on \nickname }
\subsubsection{\revAdd{Creativity Support and NASA-TLX Results.}}
The metrics we measured showed that designers responded to \nickname{} with enthusiasm. \revAdd{All responses were on a 7-point Likert scale.} In terms of enjoyment, 12/13 participants rated their experience positively ($\geq$ 5 out of 7) for $T_{edit}$ \revAdd{ (median: 6)} and 11/13 for $T_{create}$ \revAdd{ (median: 6)} . The majority of participants also responded positively that they were able to find at least one design to satisfy their goal: 10/13 respondents in $T_{edit}$ \revAdd{ (median: 6)}, 12/13 respondents in $T_{create}$ \revAdd{ (median: 7)}. 
%TC:ignore
\revDel{We note that for $T_{create}$ 7 of the participants responded with the maximum agreement (strongly agree).} 
%TC:endignore
Likewise, most participants reported that the system helped them fully explore the space of designs (9/13 responded positively for $T_{edit}$  \revAdd{ (median: 6)} , 11/13 for $T_{create}$  \revAdd{ (median: 6)}).

\begin{quotation}
\textit{``I could spend ages in this.'' - P18}
\end{quotation}

In general, the post-task questionnaire results were similar for $T_{edit}$ and $T_{create}$. However, on a few dimensions, participant responses were distributed slightly differently. For example for effort, responses for $T_{edit}$ about tool performance \textit{(``How successful were you in accomplishing what you set out to do?'')} were split across the spectrum, with 6/13 rating the tool positively  \revAdd{ (median: 4)}. For $T_{create}$, 10/13 participants rated the performance positively \revAdd{ (median: 5)}. In terms of ease of prompting, while 13/13 respondents were positive that for $T_{create}$ it was easy to come up with prompts \revAdd{ (median: 7)}, 10/13 responded positively for $T_{edit}$ \revAdd{ (median: 5)}. We hypothesize that this could have been because for $T_{edit}$ participants had to work under more constraints, bringing in 3D models that were often custom and near finished.

We note that frustration was low for both Tasks; 11/13 responded on the low side of the spectrum for $T_{edit}$ ($\leq 3$) \revAdd{ (median: 3), and} 10/13 on the low side for $T_{create}$. For $T_{edit}$ \revAdd{ (median: 2)}, frustration was low in spite of the fact that 6/13 of participants disagreed to some degree ($\leq 3$) about having control over the generations. 

\begin{quotation}
\textit{``The amount of control you have with the system is very dependent upon how specific you get with the text. For example, if I make it super broad, you're obviously going to have less control because DALL-E is working off of less information. So it may provide its own information. It has to kind of fill in the gaps of what you're trying to say. But the more specific I got, the better results I got.'' - P1}

\end{quotation}
\begin{quotation}
\textit{``It was a bit difficult to control. Some things I wasn't quite expecting. For example, with this one [generation of a watch] I expect that it would have more circular watch faces, but it came with ones that were more angular.'' - P8}
\end{quotation}

\subsubsection{\revAdd{Usefulness of GPT-3, CLIP Highlights, Image Prompts }}

Lastly, to understand how helpful different features (prompt suggestions, CLIP highlighting, automatically captured viewport images) are in the construction of text and image prompts, we discuss workflow-specific questions about the prompting pipeline of \nickname. Participants were asked about the usefulness of \nickname \revAdd{for} their usual workflow. For $T_{edit}$, 10/13 felt that it would be helpful \revAdd{(median: 5)}. For $T_{create}$, 10/13 also felt it would be helpful \revAdd{(median: 7)}.

In another question, we asked whether it was easy for participants to come up with new ways to prompt the system. Participants responded unilaterally positively for $T_{create}$ (13/13 responded $\geq 5$) and positively for $T_{edit}$ \revAdd{ (median: 6)} (10/13 responded $\geq 5$) \revAdd{ (median: 6)}. Participants were also asked to rate how useful they found the GPT-3 suggestions. For $T_{edit}$ and $T_{create}$, the responses were generally positive, at least 8/13 participants responded with 5 or higher for both tasks \revAdd{ ($T_{edit}$ median: 7, $T_{create}$ median: 6)}. 

\begin{quotation}
\textit{
\color{black}
``I'm looking for the right word and I think that's where this text [GPT-3] search can come in handy\ldots I think it's helpful to know its language, to know what it finds.'' - P4}
\end{quotation}
% ideally want to replace
\begin{quotation}
\textit{``I think having the GPT-generated ones was useful. It allowed for some ideas I didn't consider\ldots [ideas I] wouldn't have found the words for.'' - P13}
\end{quotation}
 \color{black}
On whether or not the highlighting of prompt suggestions was useful, participants responded with more even distributions, though the distributions still skewed positive (8/13 in $T_{edit}$ and 7/13 in $T_{create}$ rated the statement at 5 or higher
\revAdd{ (median: 5, for both tasks)}. \revAdd{Participants tended to click on suggestions that were highlighted more strongly for text-image alignment, often choosing the most strongly highlighted suggestion within the category.}

%TC:ignore
\revDel{
Across 570 sets of prompt suggestions from both tasks, chosen suggestions had an average highlight opacity of 0.17 (scale of 0.0-1.0); not chosen suggestions had an average highlight opacity of 0.09. We hypothesize that these averages skew low on the scale because weakly highlighted (low opacity) suggestions were also chosen when participants were interested in them, even if the text-image alignment was not high.}
%TC:endignore

Lastly, we gauged participant response to image prompts, asking if they agreed that image prompts were incorporated well in their generations. For $T_{edit}$, 10/13 participants responded with a 5 or 6 for agreement \revAdd{ (median: 6)}. For $T_{create}$, 8/13 participants responded with a 6 or 7 \revAdd{(median: 6)}. 

\begin{quotation}
\textit{
``Image prompts definitely allowed me to tailor the outcomes towards what I was hoping for or expecting maybe\ldots I'd have struggled to replicate [the render type] if I hadn't done the click on the image [sent in an image prompt] and create some variations. I think once I found something I liked, using those variations made it much easier to stuck to that design theme.'' - P13}

\end{quotation}

\begin{quotation}

\textit{``This middle one is pretty insane\ldots it has integrated my design into the image properly\ldots even as an assembly, I think that's completely nuts\ldots [An image prompt] connects what I'm working on with it [DALL-E]\ldots otherwise it might be giving some random results, and after a while it might become redundant for me.''] - P18 }
\end{quotation}

\revAdd{
We analyzed participant prompt logs to quantify how often participants used \nickname{}-provided prompt suggestions. For both $T_{edit}$ and $T_{create}$, we counted how many times participants used the \nickname{}-provided prompt suggestions (3D keywords, designs, parts, and styles) and how many times participants provided a custom keyword. Collectively, these represented all the keywords within prompts. Across both tasks and all participants, we found that \nickname{}-provided prompt suggestions accounted for 63.61\% of all prompt keywords, showing that participants heavily used the GPT-3 function of \nickname{}. We also see in Fig. \ref{fig:tcreate_promptkeywords} that \nickname{} provided the majority of prompt keywords (at least half) for 9/13 participants in $T_{create}$ and 9/13 participants in $T_{edit}$. These results are summarized in Table \ref{tab:dalle_prompt_reliance}.  }

% TC:ignore
% \revAdd{We further analyzed 570 sets of prompt suggestions  (289 for $T_{edit}$ and 281 for $T_{create}$) where participants had made a suggestion selection. We compared the average logit scores of chosen suggestions versus those of not chosen suggestions for designs, parts, and styles (the categories that had been varied in opacity). Using a t-test, we found that the average logit score of a chosen prompt suggestion was significantly higher than that of a not chosen prompt suggestion across all categories. These results presented in Table \ref{tab:clip_results} suggest that a higher logit score correlated with a prompt suggestion being chosen by a participant.}
% TC:endignore

 \revAdd{

\begin{table}[!h]
  \caption{ Source of prompt keywords across tasks, comparing the frequency of prompt keywords supplied by participants versus by \nickname{}. \nickname{} provided the majority of prompt keywords in both tasks.}
  \begin{tabular}{lll}
    \toprule 
     & Participant-provided  &  3DALL-E-provided \\
  \midrule
  $T_{edit}$& 34.95\% & 65.05\%  \\
$T_{create} $& 38.64\% &61.36\%\\
   Both tasks & 36.39\% & 63.61\%\\

  %    \color{blue} $T_{edit}$& \color{blue} 34.95\% & \color{blue} 65.05\%  \\
  % \color{blue}$T_{create} $&\color{blue} 38.64\% &\color{blue} 61.36\%\\
  %  \color{blue}Both tasks & \color{blue}36.39\% & \color{blue}63.61\%\\

  \bottomrule
\end{tabular}
%\newline
%\\[0.5em] 
%  \revAdd{ Frequency of prompt keywords supplied by participant or \nickname{}. \nickname provided the majority in both tasks. }
  \Description{

  }
  \label{tab:dalle_prompt_reliance}
\end{table}
}
%TC:ignore
% \begin{table}[!h]
%   \caption{ \revAdd{CLIP Scores for Chosen and Not Chosen Suggestions by Categories }   }
%   \Description{

%   }
%   \begin{tabular}{llll} 
%     \toprule
    
%      & \color{blue}\textbf{Design Avg}  & \color{blue}\textbf{Part Avg}  & \color{blue} \textbf{Style Avg} \\
% \color{blue} Not Chosen Suggestion & \color{blue} 0.095 &\color{blue} 0.082& \color{blue} 0.092  \\
%  \color{blue} Chosen Suggestion & \color{blue} 0.163 & \color{blue}0.182 & \color{blue} 0.175 \color{black}\\
%  \color{blue} p-value & \color{blue} 0.02 &  \color{blue} <0.01 &  \color{blue}0.02 \\
%   \bottomrule
% \end{tabular}
% \\[0.5em]
% \revAdd{
% Average logit scores of chosen vs. not chosen prompt suggestions for design, part, or style categories. Chosen prompt suggestions were significant higher in average logit score than not chosen ones.}
%   \label{tab:clip_results}
  
% \end{table}
% \color{black}
%TC:endignore

\begin{figure*}
    \centering
    \includegraphics[width=0.85\textwidth]{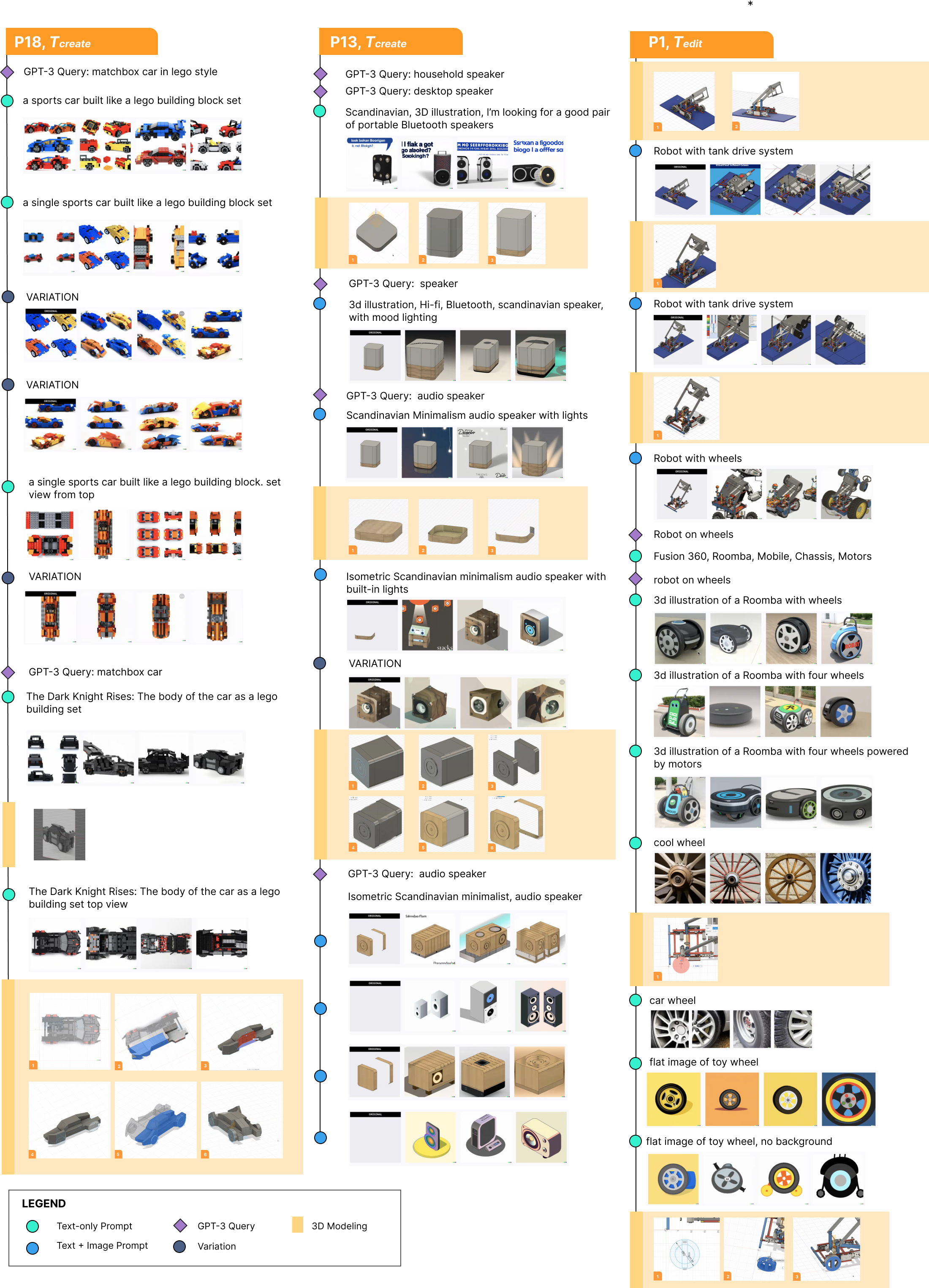}
    \caption{\revAdd{Prompting and 3D modeling workflows of design process of three participants (P18, P13, and P1). P18 created a car, P13 created an audio speaker, and P1 edited a robot. Timelines are vertical with the markers representing different generation requests and yellow intervals representing CAD time. The markers preserve order but the time stamps across participants are not aligned / to scale.} }
    \Description{Prompting and 3D modeling workflows of design process of three participants (P1, P13, and P18). P18 created a car, P13 created an audio speaker, and P1 edited a robot. Timelines are vertical with the markers representing different generation requests and yellow intervals representing CAD time. The markers preserve order but the time stamps across participants are not aligned / to scale.}
    \label{fig:example_reference}
\end{figure*}

\subsection{Prompting Behavior}

\begin{figure*}
    \centering
    \includegraphics[width=\textwidth]{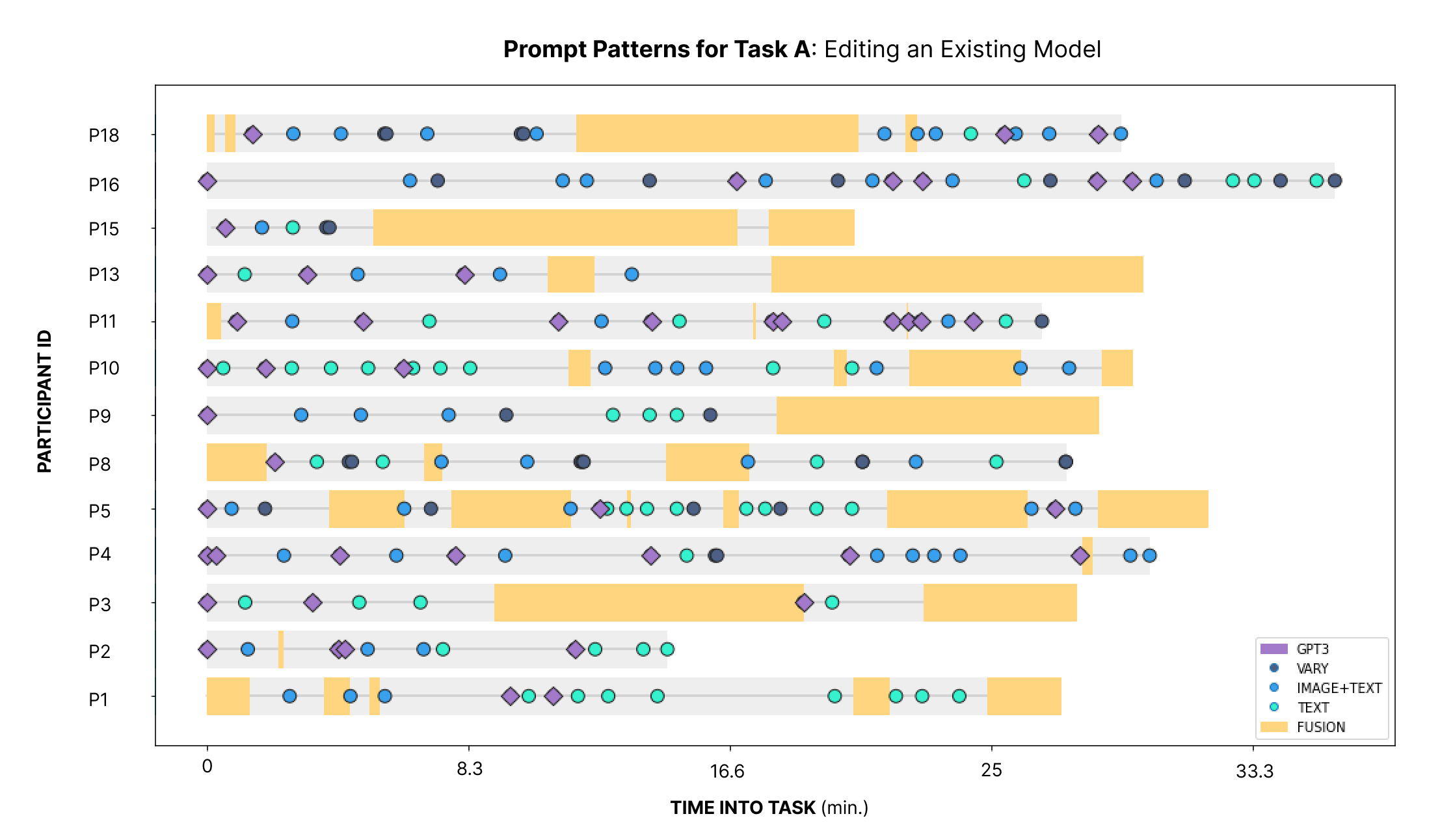}
    \caption{Pattern of generation activity for $T_{edit}$, when participants edited an existing model.}
    \Description {
    Thirteen row horizontal bar chart plotting out generation activity for T-edit. Purple diamonds marked GPT-3 requests, while circle dots marked DALL-E requests. Backgrounding the markers were time spent in Fusion (colored orange intervals) versus time spent in 3DALL-E. Patterns are elaborated and interpreted under Evaluation.
    }
    \label{fig:patterns_a}
\end{figure*}

\begin{figure*}
    \centering
    \includegraphics[width=\textwidth]{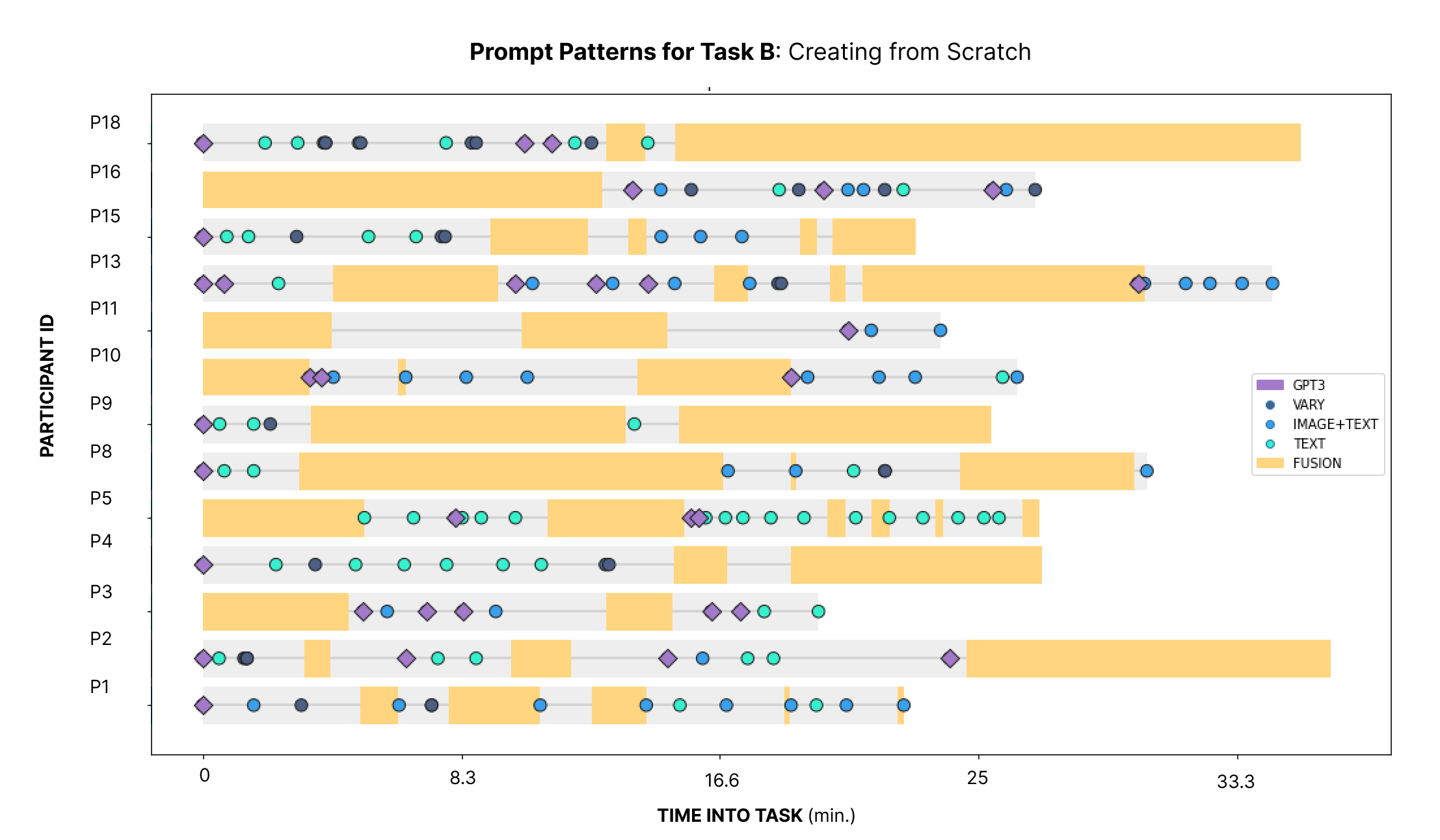}
    \caption{Pattern of generation activity for $T_{create}$, when participants created a model from scratch.}
     \Description {
    Thirteen row horizontal bar chart plotting out generation activity for T-create. Purple diamonds marked GPT-3 requests, while circle dots marked DALL-E requests. Backgrounding the markers were time spent in Fusion (colored orange intervals) versus time spent in 3DALL-E. Patterns are elaborated and interpreted under Evaluation.
    }
    \label{fig:patterns_b}
\end{figure*}

We were able to observe certain patterns of prompting with \nickname{} as each generation action was logged by our interface. From these logs for both GPT-3 and DALL-E, we were able to provide timelines of generation activity in Fig.~\ref{fig:patterns_a} ($T_{edit}$) and Fig.~\ref{fig:patterns_b} ($T_{create}$).

\subsubsection{AI-first, AI-throughout, or AI-last.} One of the most salient ways to distinguish participants was at which points in their workflow they took to \nickname{} and \revAdd{at} which points they focused on Fusion 360. Some participants were \textit{AI-first}, meaning they tended to sift through AI generations first until they had a better grasp of its abilities or until they found a design that they liked before taking any significant 3D design actions. For example, P18 (top row of Fig.~\ref{fig:patterns_b}), a technical software specialist with an industrial design background, was trying to make a car. They first began looking for inspiration for a matchbox car, before diving into prompt suggestions like ``sports car''. Text prompts that P18 tried included \textit{``a single sports car built like a Lego building block, view from the top.''} and \textit{``The Dark Knight Rises: the body of a car as a Lego building set''}. \revAdd{ They added perspective (``view from the top'') and a number word (``single'') to specify the composition of their generation and tried ``The Dark Knight Rises'' as a style suggested by \nickname{} for the query 
\textit{``matchbox car''}. } After liking one of the resulting generations (Fig.~\ref{fig:examples_reference2}), P18 used the result as a reference image. For the rest of the duration of the task, P18 modelled within Fusion 360. \revAdd{ 
  P18 first traced over half of the generation like a blueprint before extruding faces to varying heights. They then beveled and chamfered these starting blocks of a car to add ridges and windshields and subtracted material to make room for wheels. They ended by mirroring the half of the car they modeled to create a full symmetrical car.} 
  %TC:ignore
  \revDel{ The modeling experience during $T_{create}$ is shown in Fig.~\ref{fig:example_reference}.} 
  %TC:endignore
  \revAdd{P18's prompting and modeling workflow for $T_{create}$ is shown in Fig.~\ref{fig:example_reference}.}

The \textit{AI-last} pattern occurred when participants jumped straight into their existing workflows for 3D design and tried \nickname{} later. We see this in the rows of Fig.~\ref{fig:patterns_b} that start off with orange bars, which indicate that participants started modeling from the get-go of the task. P11, for example, was trying to make a bottle. They began by sketching the cross-section of a bottle and revolving it 360 degrees to create a form. After filleting the base to round it and hollowing it out with a hole, they found prompt suggestions from \nickname{} like \textit{``fusion 360''} and \textit{``Coca-Cola''}. Using a generation prompted from \textit{``Front view Coca-Cola Bottle''}, they edited their bottle cross-section to match that of the generation.
Only after they had created this basic bottle did they start looking for inspiration; seeing generations of Coca-Cola bottles \textit{later} helped P11 figure out how to bring complexity into the cross-section of their design.
P16 (second row in Fig.~\ref{fig:patterns_b}) was another AI-last participant. They already had an existing screwdriver concept in their mind. \revAdd{ They began by sketching and extruding a rounded rectangle for the grip of the screwdriver, dimensioning accordingly. They worked on the flat-head tip by extruding a narrow cylinder and lofting the face out to a point. After making a rough model, they tried \nickname{} with prompts specific to flat-head screwdrivers and used their existing modeling progress as an image prompt. P16 commented that \nickname{} inspired them to consider different handle cross-sections (e.g. hexagonal, square) and grooved grips}. Note that the AI-last pattern, jumping into a participant's existing workflow with Fusion 360, was more prevalent in $T_{create}$.

However, there were also participants who queried \textit{AI-throughout}. Many participants (P13, P1, P8, P10) would intermittently craft an image prompt by briefly working within Fusion 360 and then start generating. We see these actions whenever participants would have a short window of Fusion time that led up to image+text generation (medium blue dots in Fig.~\ref{fig:patterns_a} and Fig.~\ref{fig:patterns_b}). During these short windows, participants were generally changing their camera perspective or the visibility of different parts in their assemblies. For example, P10 hid the hopper of a toy truck they had brought in and tried to generate different semi-trailers using prompts such as \textit{``Jeep Gladiator snow plow truck''}. 
%TC:ignore
\revDel{P13 went a step further and performed destructive operations on their model geometry, deleting faces and extrusions in their geometry to get the right image prompt they wanted. }
%TC:endignore
\revAdd{P13 (during $T_{create}$) was another \textit{AI-throughout} participant. They first built up a base for an audio speaker they wanted to design and applied wood and chrome finishes for a Scandinavian design aesthetic. They then tried prompts with lighting elements  (e.g. \textit{``Isometric Scandinavian minimalism audio speaker with built-in lights''}). They built towards a generation they liked for a while, adding details of a speaker cone and applying tessellation and reducing operations to give the speaker body structural texture. Then they began to create image prompts for \nickname{} to fill in---deleting faces and extrusions or hiding bodies in their geometry. They wanted to see the different ways the middle section of their speaker could be autocompleted. We see P13's work in $T_{create}$ and the way they utilized AI-throughout their workflow illustrated in Fig. ~\ref{fig:example_reference}. }

Participants would also use text-only prompts to take them towards new directions. P9 used text prompts to pivot their design multiple times and better scope their 3D design. Originally, P9 intended on creating a prosthetic hand and tried generating \textit{``A 3D model of a robotic hand with two fingers''}. After finding modeling a hand to be too complex because of how articulated they are, they tried text-only prompts \textit{``3d model of a human fist''} and \textit{``3d model of mittens''} to explore what they could more feasibly model, exploring divergently. Deciding on mittens, they imported a generation as a reference and sketched over it. After extruding the sketch and applying fillet operations to round out the mittens, P9 added cuff sleeves, a detail inspired by the generation.

In terms of generation patterns for GPT-3, nearly everyone started with generating from GPT-3 (though this could be because of the organization of the user interface). Many continued to use GPT-3 throughout each task, and we can see this reflected in the fact that there are purple diamonds (GPT-3 actions) at the early, middle, and late stages of workflows for both $T_{edit}$ and $T_{create}$.

\subsubsection{Switches between Types of Prompting}
Eight participants passed in \textit{an image prompt} as their first action in $T_{edit}$, and eight participants passed in \textit{text prompts} as their first generation action for $T_{create}$. This suggests that participants may be more likely to pass in an image prompt if they already have work on their page. Aggregating across all the different generations across $T_{edit}$ and $T_{create}$, we did not see that any mode of prompting was favored more than the rest. Preferences in prompting were highly dependent upon the participant and also how well the participant felt like the generations incorporated their image prompts. For example, even though P13 found image prompts useful, they felt like image prompts were incorporated in an ``awkward'' way, as they had more glaring visual artifacts than text-only generations. %TC:ignore 
\revDel{Others like P4 were delighted by the way the visual artifacts could take their models in unexpected directions. For example, they greatly enjoyed how DALL-E produced a kaleidoscopic image out of their machined part.}
%TC:endignore

In certain rows in Fig.~\ref{fig:patterns_a} and Fig.~\ref{fig:patterns_b}, we could see that some participants would shift away from using image prompts and focus on text-only prompts. A case in point of this was when P1 worked on a tank-drive robot \revAdd{that they had built for a FIRST \cite{first_2022} robotics competition} during $T_{edit}$ \revAdd{(pictured in Fig. ~\ref{fig:example_reference})}. To craft image prompts, they played around with different angles of their models and toggled the visibility of parts like the wheels and ground plane of their model. \revAdd{The robot was a highly convoluted assembly, and} while they found that 3DALL-E could generate decently even on these visually complex image prompts, they ended up passing in a series of text-only prompts like \textit{``3D illustration of a Roomba with four wheels powered by motors''} and \textit{``flat image of a toy wheel''} (focusing in on a specific part rather than trying to get \nickname{} to work with the full assembly was also a common strategy of participants\revDel{.})\revAdd{.} In this situation, the text-only generations were easier for P1 to parse and make sense of. P5 was another example of someone who pivoted away from passing in image prompts to use text-only prompts after receiving sets of unsatisfying generations during $T_{edit}$. \revAdd{ The image prompt that they passed in was a mechanical base, so the generations building off of that were all visually indeterminate (not recognizable as any particular object).} P5 instead decided to generate textures of water and maple syrup to project onto their original model (as seen in Fig. ~\ref{fig:examples_b}), finding this to be an easier way to make use of their part and \nickname{}. 

%TC:ignore
\revDel{People also tended to utilize text-only prompts when they wanted to pivot design directions.P8 used text-only prompts to try different design styles (from ``industrial minimalist'' to ``traditional farmhouse'' to ``nature-inspired Scandinavian'') in quick succession. }
%TC:endignore

\begin{figure*}
    \centering
    \includegraphics[width=0.8\textwidth]{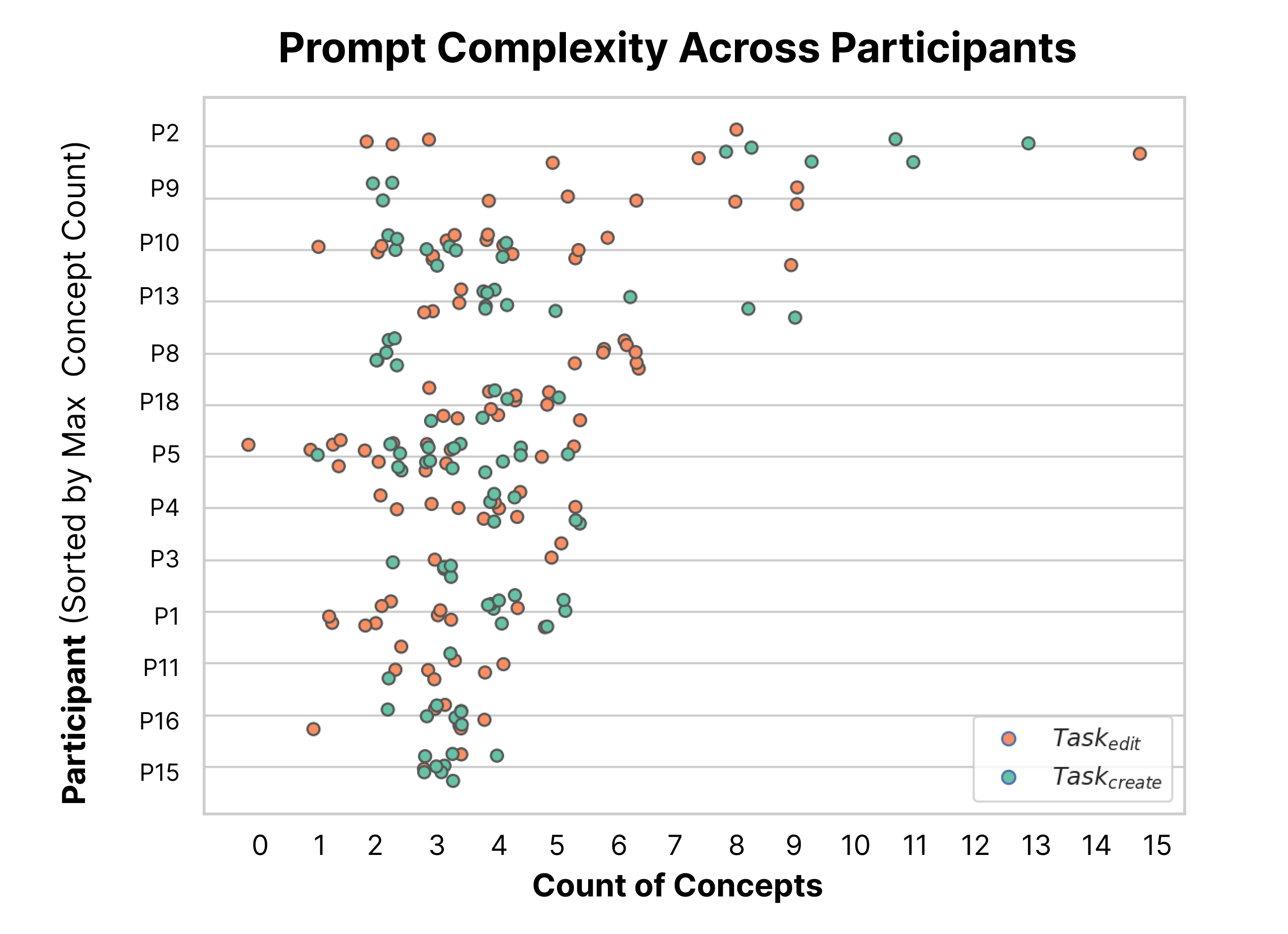}
    \caption{Prompt complexity measured across participants, where complexity is the count of concepts in each text-only and image+text prompt. Participants span the X-axis, sorted by the count of their most complex prompt. The values are jittered to show multiplicity; many prompts mapped to the same number of concepts.  Complexity tended to concentrate between two to six concepts, as seen by the density of prompts within that interval. Each datapoint was colored based on prompt task. }
    \Description{Scatterplot resembling a strip plot of count of concepts in each prompt measured against participants on y-axis. Generally, data points concentrated between two to six concepts, with one outlier on the line x=2 (for the second participant)  }
    \label{fig:prompt_complexity}
\end{figure*}

\section{Prompt Complexity}
% \revAdd{Figure moved to supplementary as well}
\revAdd{It can be challenging for an end user to understand how lengthy or detailed a text-to-image prompt should generally be, which is why we studied prompt complexity with \nickname{}. }
In \nickname, GPT-3 would automatically rephrase selected prompt suggestions while adding a small amount of connecting words. Based on this design, we could measure complexity as the number of concepts forming the basis of a prompt. For example, if \textit{``3d render, minimalist, chair''} was rephrased as \textit{``3d render of a minimalist chair''}, we gave the prompt a count of 3 concepts.

However, participants also had the ability to edit the final prompt and to add or subtract concepts of their own. In cases where the text prompt mostly came from the participant rather than GPT-3, we counted the number of concepts based on rules from linguistics and natural language processing. The prompt complexity was then the number of noun phrases and verbs in a prompt, ignoring prepositions, function words, and stop words. Count words were ignored; they were considered modifiers for the noun phrases they were a part of (e.g. ``five fingers'' was one concept).

We annotated text-only and image+text prompts with the number of concepts. We did not annotate variations for complexity because the generation of those images were not directly informed by text prompts. From these annotations, we charted prompt complexity across participants in Fig.~\ref{fig:prompt_complexity}. We found that participants tended to explore between two to six prompts, which is where most of the density of points concentrates in Fig.~\ref{fig:prompt_complexity}. We see that participants were also willing to try a range of concepts, as we can see in the wide spread of P2, P9, and P10. Fig.~\ref{fig:prompt_complexity} also shows that participants could easily assemble prompts of over six concepts with this workflow.

We note that even when the prompts were filled with concepts: \textit{``V-shape,Y, Tricopter, Sports, Abstract, Landscape, Aerial, Gimbal, Camera, Transmitter, Flight controller, Receiver''} , \nickname{} could still return legible images. For this prompt, P2 received generations that had laid out displays of product components. P2 was an obvious outlier in the complexity of the prompts that they provided. They were keen on trying to ``break the system'' and passed prompts averaging 10 concepts. We did not discern a difference between complexity observed for $T_{edit}$ and $T_{create}$.

\section{Qualitative Feedback}

% \subsection{\sout{Industry Use Cases}}
\revAdd{\subsection{\nickname{} Use Cases for CAD Design}}

\subsubsection{Use Case: Preventing Design Fixation}

Participants demonstrated different use cases of \nickname{} as they progressed through the tasks. The most commonly acknowledged use case was using the system for inspiration, particularly in the early stages of a design workflow. P10 contextualized some of the challenges that 3D designers face on the job, such as design fixation and time constraints. \textit{``A lot of times designers get stuck, they get tunnel vision...the folks at [toy design company] used to say to me, ``We can't come up with enough designs...it takes too long to come up with a design, so then we only get two or three...we would like to see thousands of design options and variations...the [designer's] goal is to start throwing as many designs out there as they can.''} 

\color{black}

Participants felt like the tool could be \textit{``game-changing''} (P11) for certain industries such as consumer products, automobiles, and game assets (P10, P11, P15, P17, and P18). They likened it to existing search and intelligent suggestion tools like stock photography websites (P15) and Google Images, but noted that with \nickname, it was better in that users could access inspiration without leaving their workbench (P11). \revAdd{ For example, P8 was building a table and explored many different design styles from \textit{``industrial minimalist''} to \textit{``nature-inspired Scandinavian''} by using text-only prompts in quick succession. } They also passed in the table top they had already modelled to see how it could be completed by \nickname{} as a \textit{``CGI traditional farmhouse table with centerpiece drawers``}. \revAdd{They did all of this without having to switch applications, which is important for 3D design software, as it requires focused modeling time.} %TC:ignore
\revDel{The lack of application switching is important in 3D design software, which requires focused modeling time. In the following sections, we elaborate use cases for text-to-image AI in 3D workflows.}
%TC:endignore
\begin{figure}
    \centering
    \includegraphics[width=0.47\textwidth]{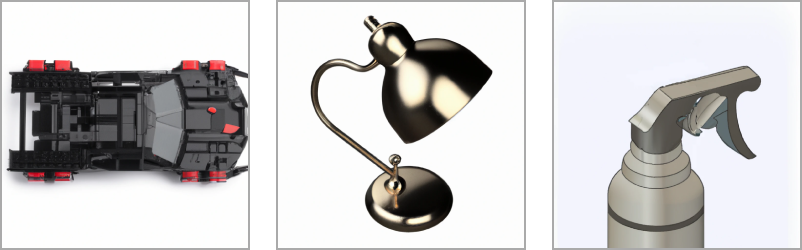}
    \caption{Three DALL-E generations participants (P18, P15, P9) found inspirational from the prompts: \textit{``The Dark Knight Rises: the body of a car as a Lego building set top view''}, \textit{``3D render of a desk lamp Victorian''}, and \textit{``isometric 3d renders of a cleaning sprayer bottle''}.}
    \Description{Three DALL-E generations participants (P18, P15, P9) found inspirational from the prompts: "The Dark Knight Rises: the body of a car as a Lego building set top view", "3D render of a desk lamp Victorian", and "isometric 3d renders of a cleaning sprayer bottle".
    }
    \label{fig:examples_reference2}
\end{figure}

\subsubsection{Use Case: Reference Images for 3D Geometry}

Many participants (P18, P3, P13, P11) imported generations into the 3D software as reference images to model off of. P18 and P13, both of whom had backgrounds in industrial design, described how designers traditionally gather reference images to build their models as part of their CAD workflows. These images generally aligned with specific views: front, side, top-down, perspective, or isometric. P18 said, \textit{``I would probably need at least three images: top, side, and front view to even understand it three-dimensionally...that's what a designer would pass to an engineer to then build it. I would try to force it [\nickname] to create a top view, side, front view that are somehow matching.''}
 %TC:ignore
 \revDel{After generating a top view image, P18 was able to import that image as a canvas and construct off of it. The process, where P18 sketched and extruded geometries upwards from a plane parallel to their reference image, is pictured in Fig.~\ref{fig:example_reference}.However, } %TC:endignore
 
 \revAdd{P18 used a top view generation as a basis for their model as shown in Fig. \ref{fig:example_reference}. } We note that most generations came back angled and at perspectives unless the prompt explicitly specified viewpoints like ``top view'' or ``flat'', and that \nickname{} did not always capture ``isometric'' and ``perspective'' views in the technically accurate sense of those words. Nonetheless, even if generations were not drawn to perspective or as clean as technical drawings and renders usually are, participants still found \revAdd{them useful as reference images}.

Other participants used the generations as references albeit more loosely. P15, liking a \textit{``3D render of a desk lamp Victorian''} (Fig.~\ref{fig:examples_reference2}), made the arm of their lamp skinnier as per the generation. 
P9, observing generations from prompts such as \textit{``isometric 3d renders of a cleaning sprayer bottle''} (Fig.~\ref{fig:examples_reference2}), noted that they could subtract volume from the outer contours of their model and reduce the amount of material used, which was part of their goal to design a more sustainable spray bottle top.

\subsubsection{Use Case: Textures and Renders for Editing Appearance}
Participants would also edit their model appearance towards the look of generations (P13, P15, P10, P3). They could do this by applying textures within the software and dragging and dropping materials from the software's material library onto surfaces. 
%TC:ignore
\revDel{For example, P13 dragged wood and chrome finishes onto their speakers model and tried texturizing the faces of their model with tessellation to match the color palette and low-poly look of a generation.} 
%TC:endignore
P5, innovatively used generations as textures to help build a 3D outdoor movie theater scene. Their scene was built out of simple geometries, and atop these geometries, they placed generations of a \textit{``jello bed''} and generated portraits of pop culture characters (pictured in Fig. \ref{fig:examples_b}).

P1 mentioned that \nickname{} could be useful for product design presentations to show the function or interaction of things being designed. As P1 made a prosthetic hand, they imported a generation and started to model atop it. Curious about how a text+image prompt would fare if it included a generation transparently overlaid over their geometry, they generated and found compelling images \revAdd{that could visually situate their designs with their use cases in product design presentations}.

\subsubsection{Use Case: Inspiring Collaboration}

Design in industry is a team effort, and while \nickname{} was evaluated in the context of a single user, many participants acknowledged that \nickname{} could be beneficial in teams. P16 mentioned that from their industry experience, \nickname{} would be excellent for establishing communication between mechanical engineers and industrial designers. Mechanical engineers focus on function, while industrial designers focus on aesthetics. P16 felt that \nickname{} could help both sides pass around design materials for discussion and common ground.

P13, who was an industrial designer, noted that teams could also do multi-pronged exploration with \nickname. Because each team member would have individual prompting trajectories, a team could easily produce diverse searches and more variety during brainstorming. P3 mentioned that there are already points within their industry (automotives)  where there are hand-offs between the people who generate design ideas and the people who execute them. Technical sales specialist P4 also mentioned that they could instantly see \nickname{} being useful for their clients, many of whom have bespoke requests such as organic fixtures for restaurants and museums or optimized shapes for certain materials. 
%TC:ignore
\revDel{While they did not see the tool as a problem solver itself, they felt it could provide ideas that would help them do the problem solving parts of their job that they love doing.}
%TC:endignore
\subsubsection{Use Case: Inspiring Design Considerations}

\nickname{} also inspired design considerations by making participants think about different aspects such as functionality or manufacturability. For example, P1 was looking for a wheeled robot. Seeing generations where robot bodies were varied in the number of wheels they had or how far off the ground they were made P1 think about the different amounts of motor power these robots would require. While \nickname{} could not guarantee the feasibility of every generated design, some participants (P1, P8) liked that \nickname{} inspired them to think through details such as how manufacturable a design was. 

Participants also felt like they could elicit unique, out-of-the-norm designs from \nickname{} and use it to \revAdd{let them} gauge the uniqueness of their own designs. P4 wanted to design a product that did not exist in the real world yet: an ear gauge electronic for their son. They treated the model's inability to come up with their exact vision in generations as a good thing, interpreting it to mean that the product did not exist yet and therefore had patentable value. \textit{``We [DALL-E] started to lose a little bit when we started putting in the `Bluetooth ring', which is good because that tells me\ldots probably out there in the real world, nobody's actually doing this\ldots that made me feel good about the fact that I might have a predicate design in my head.''} P2, who had taught drone design classes, also felt like right off the the bat, \nickname{} was able to produce unique aesthetics beyond what is typically seen in drones, something their students generally struggled to do. P15 also felt like \nickname{} could have educational value as they looked around for ways to accomplish something they saw in a lamp generation: \textit{``being able to reverse engineer\ldots that is a cool learning aspect.''} 3DALL-E could not guarantee the educational or patentable value of a generation, but it inspired participants (P4, P2, P15) to think about design considerations such as design conventions, uniqueness, and plausibility.

\subsubsection{Weaknesses in terms of CAD}
Some participants did comment that text-to-image AI may have weaknesses in applications like machining and simulation or the construction of internal components and other function-focused parts. P9 pointed out that it would be difficult to generate geometries that enclose parts, because if a user was to pass in an image prompt of that part, \nickname{} would be unable to draw housing over it. Likewise, a participant mentioned that they could imagine \nickname{} being used to design the facade of the car, but they did not believe that it could design a more internal component not easily describable in layman's terms. 

% \begin{figure*}
%     \centering
%     \includegraphics[width=\textwidth]{fig/prompt_bibliography_final.png}
%     \caption{Prompt bibliographies, a design concept we propose for tracking human-AI design history. As prompts become a part of creative workflows, they may be integrated into the design histories already kept by creative authoring software. This bibliography tracks text and image prompts, as well as which generations inspired users during the tasks.}
%     \Description{
%     Two columns show concept of prompt bibliographies, which are rows of generations. There are 6 rows of 4 generations each for a total of 24 images in each column. There are stars on generations that were favorited by participants. Each row is labeled by the text prompt and image prompt if any. Left column is of generations for a speaker done by P13. Right column is of generations for a robot by P1.
%     }
%     \label{fig:prompt_bibliography}
% \end{figure*}

\subsection{\revDel{Incorporation into Diverse Workflows} \revAdd{ Comparing with Traditional Workflows}}

Our exploratory study invited designers to stress test \nickname{} across the settings of a wide range of disciplines. Participants were impressed with the ability of the model to generate even when they passed prompts filled with technical jargon like ``CNC machines'', ``L-brackets'', or ``drone landing gear''. Still, prompting remains very distinct from the workflows participants usually go through. Many participants described their regular design process as multiple phase progressions from low fidelity to high fidelity. They mentioned roughing out designs first, putting placeholders within robotic assemblies (P1), box blocking up to complexity (P13), and redesigning from the ground up again and again (P18). Even though \nickname{} only provided images of 3D designs, these designs could have high fidelity details that could shortcut participants to later stages of the design process. 

\subsubsection{Text Interactions in 3D Workflow}
The most distinct difference in workflows is that \nickname{} is text-focused, but text is not central to 3D design workflows, which are usually based on the direct manipulation of the geometry. P13 mentioned that designers primarily operate visually. \textit{``The only reason I really use text in an industrial design context is [for] making notations on a design...to explain what a feature is...to write a design specification...but the majority of the time is image focused.''} Because of this, P13 preferred the ``image-based approach'' within \nickname{} where they could ``provide it with a starting image and get variants of that''. P4, however, thought that in some respects designers \textit{are} often engaging with text, but in the form of numbers, properties, parameters, equations, and configurations. \textit{``[We] do it in a smart way...[we] drive it with the math equation. This is something we can do in parameters, and it is very text-based.''}

\subsubsection{Problem Solving with \nickname}
P10 and P4 described their day-to-day job tasks as customer-facing CAD specialists as problem solving and finding design solutions. P10 began the study wondering if \nickname{} could solve a problem they were facing in their job: packaging a toy truck. To do so, they like many of the participants, tried employing \nickname{} as a problem solver. P10 tested prompts such as \textit{``create a toy dump truck and fire truck with plastic material''} and \textit{``protect a sphere with foam''} to see if \nickname{} could help encase a 3D model. From the results they saw, they concluded that \nickname{} \textit{``was not intended to be a problem solver type of tool''}.

P13 set up image prompts as autocompletion problems. As they built an audio speaker for $T_{create}$, they commented that they were \textit{``creating two pieces of geometry and using it [\nickname{}] as a connection between the two\ldots kind of like the automated modeling command''}~\cite{automated_modeling}. They also tried other innovative ways of creating image prompts: \textit{''a hacky approach, trying to keep preserved geometries with the faces and using \nickname{} to fill in the gaps''}. %TC:ignore
\revDel{ P13 also tried other complex image prompts by deleting faces and hiding parts of the geometry to create transparent places for DALL-E to fill in, even if decimating parts of the geometry would not be a natural part of their workflow.}
%TC:endignore

\subsubsection{Driving the Design}
When AI input is added into a workflow, questions of who drives the design process and who owns the final design can arise. While P9 liked that \nickname{} augmented their workflow with what they called dynamic feedback, they felt as though their design was being driven by the generations. \textit{``Initially, the image did not really meet my expectation\ldots but eventually I was also trying to not imagine anything and just depend upon what it was suggesting.''} P3 mentioned that they felt as if they were driven by \nickname{}, while P15 mentioned that sometimes in the midst of exploring, they felt they were not gravitating towards building.

As for ownership, many participants felt like the designs they created with \nickname{} would still be their own. P1 stated on ownership, \textit{``A lot of 3D modeling is stealing...borrowing premade files online, and then assembling it together into a new thing. For this robot, we borrowed these assemblies from already premade files that were sold by the company. We modelled based off of that, but the majority of this robot can be considered ours because we determined the placement.''} P13 was also not worried about ownership concerns, stating that even now, anyone can recreate any model found online, but that \textit{``it's about the steps you go through to get there.''} 

P18 mentioned that for an AI to be applied to the real world, it still takes an expert designer's understanding of the market and customer needs. \textit{``I would use my know-how of manufacturing processes and the market or style. My service would adopt AI as a source of inspiration rather than as the solution.''} Reflecting on if AI inspiration became mainstream without designers in the loop, they expressed concerns that \textit{``if everyone would converge on the same designs [because] it only learns from the input it gets from people\ldots we might lose creativity.''}
 
\subsection{Comparison with Existing Generative CAD Tools}
Five of 13 participants had experience with the existing generative design mode within the 3D CAD software \cite{autodesk_gd_2022}. Generative design (GD) is an environment in Fusion 360 in which the completion of a 3D design is set up like a problem: users define physical constraints and geometric filters that allow a model to be autocompleted. We did not directly compare with \gdnickname, because hardware constraints made \nickname{} incompatible with \gdnickname. However, we did ask participants with experience in \gdnickname{} to compare and contrast the two.

A primary difference was that \gdnickname{} allows users to directly manipulate the model geometry, which differs from the text-based interaction of \nickname{}. \gdnickname{} results therefore free the user from doing more modeling work. What one participant liked about \gdnickname{} was that \textit{``once they set up the problem, they could just hit go\ldots don't have to actually worry about lofting and modeling''}. However, participants mentioned that \gdnickname{} has a higher barrier of entry; users are burdened with calculating loads and non-conflicting constraints, which requires some understanding of physics and engineering. 

\begin{quotation}
\textit{``You're [\gdnickname] focused on strength, endurability of the model itself, really driven as a manufacturing task\ldots your end result is something that's makeable\ldots whereas this process [\nickname] is more on the creative side.''}
\end{quotation} 

P2 mentioned that \nickname{} allowed users to come up with outcomes far more efficiently than \gdnickname. In the span of a 30-minute task, users were able to browse hundreds of results, with the first results coming in a matter of seconds, whereas P2 has previously had to wait multiple hours or even days for GD. P2 and P18 were enthusiastic that GD and \nickname{} could merge. P10 suggested that one way these two tools could complement each other is if \textit{``this tool [\nickname{}] could be used to generate shapes\ldots pass it off to the generative design [\gdnickname{}] to optimize''}.

\section{Discussion}
% \nickname{} provides the ability for to co-design conceptually with text-to-image AI is a novel capability not seen in CAD software.

%TC:ignore
\revDel{ Our results demonstrate enthusiasm for text-to-image tools within 3D workflows. With \nickname, people had a tool to help them combat design fixation and get a variety of inspiration. Furthermore, we elaborated different patterns of prompting that can identify endpoints at which text-to-image can help. 
In measuring prompt complexity, we showed that many prompts fall within a range of two to six concepts.We were also able to capture participant intentions and rich design history in the form of prompt bibliographies, as shown in Fig. \ref{fig:prompt_bibliography}. } 
%TC:endignore
\revAdd{Our results demonstrate high enthusiasm for text-to-image tools within 3D workflows. With \nickname, participants had a tool for conceptual CAD that could help them combat design fixation and get a variety of reference images and inspiration. Furthermore, we elaborated prompting patterns that can help understand when and what types of text-to-image generation can be most helpful. In measuring prompt complexity, we showed that many prompts fall within a range of two to six concepts, providing a heuristic that can be implemented in text-to-image prompt interfaces.}
 The following discussion focuses on best practices for helping 3D designers bring their own work into AI-assisted design workflows and the implications of these workflows.

\subsection{Prompt Bibliographies}

A strength of studying 3D workflows was that there was no conflict between the AI and human on the canvas, as the AI had no part in the physical realization of the design. We believe this helps mitigate ownership concerns and makes text-to-image AI very promising for 3D design tools. 
\revAdd{Currently, AI-generated content is a gray area due to concerns of attribution and intellectual property~\cite{the_new_york_times_2022}}. \revAdd{Currently, there} is no way to tell how heavily an AI-generated image borrows from existing materials. %TC:ignore 
\revDel{ It is difficult to embed signatures within AI-generated images. DALL-E generations come with small watermarks at the bottom of each image, though these can be easily cropped out.} %TC:endignore 
As generated content becomes more prevalent on platforms, it is important to develop practices of data provenance~\cite{provenance}.
We propose the notion of \textit{prompt bibliographies} to provide information on what informed designs and to separate out which contributions were human and which were AI. These can work to clarify ownership and intellectual property concerns.

\begin{figure*}
    \centering
    \includegraphics[width=\textwidth]{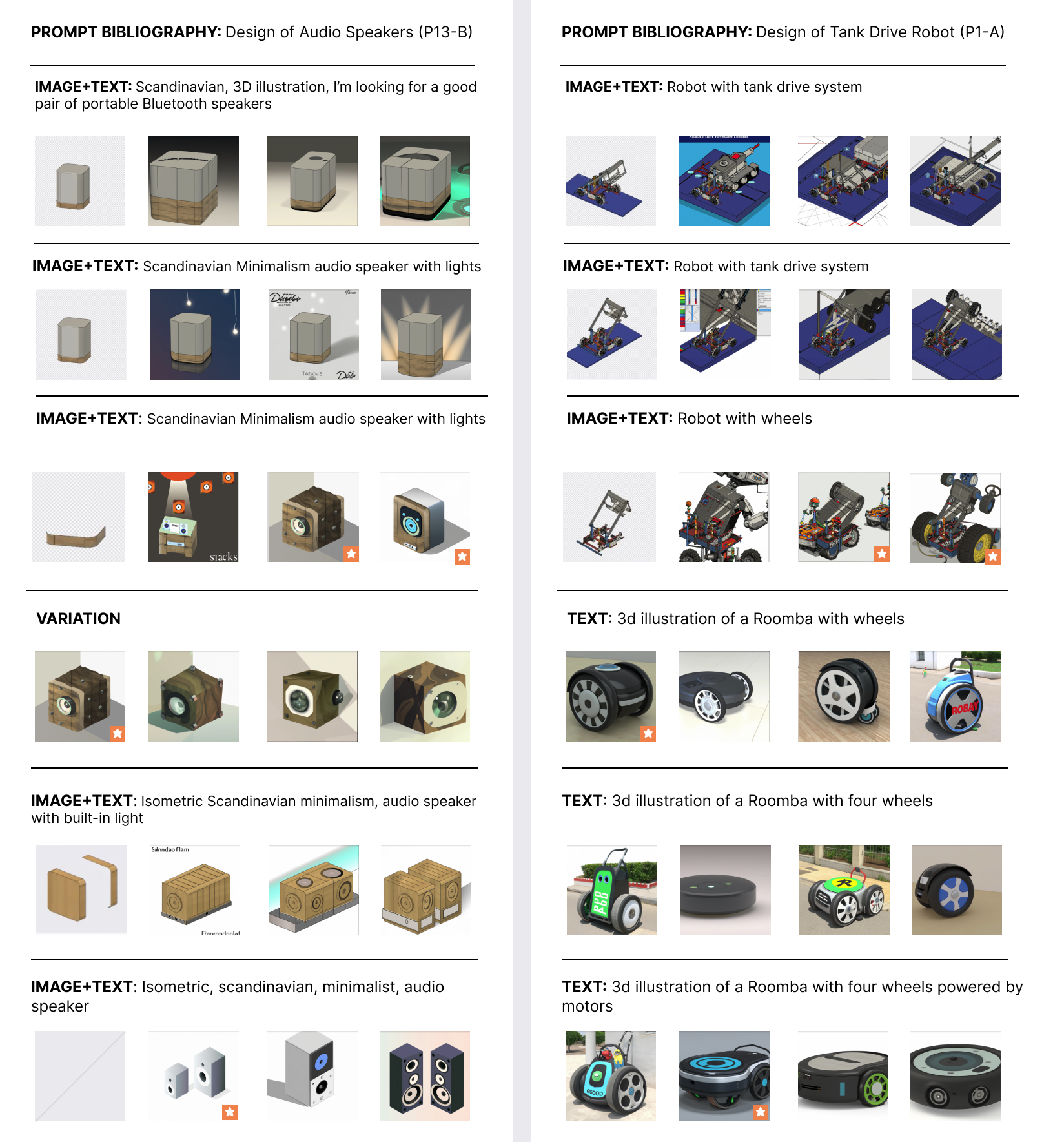}
    \caption{Prompt bibliographies, a design concept we propose for tracking human-AI design history. As prompts become a part of creative workflows, they may be integrated into the design histories already kept by creative authoring software. This bibliography tracks text and image prompts, as well as which generations inspired users during the tasks.}
    \Description{
    Two columns show concept of prompt bibliographies, which are rows of generations. There are 6 rows of 4 generations each for a total of 24 images in each column. There are stars on generations that were favorited by participants. Each row is labeled by the text prompt and image prompt if any. Left column is of generations for a speaker done by P13. Right column is of generations for a robot by P1.
    }
    \label{fig:prompt_bibliography}
\end{figure*}

Prompt bibliographies, illustrated in Fig.~\ref{fig:prompt_bibliography}, could likewise help track designer intentions and enrich the design histories that software tools provide, which generally capture commands and actions (but not intentions). The bibliographies can be merged within the history timeline features that are present in tools like Fusion 360 and Photoshop, helping prompting integrate better with the traditional workspaces of creative tools.

Sharing prompt bibliographies with their outcomes (i.e. 3D models) can also help respect all the parties that are behind these AI systems. End users can easily query for the styles of artists (as they already do) and create derivative works that dilute the pool of images attributed to artists. Prompt bibliographies may be especially relevant for CAD designers as CAD is highly intertwined with patents, manufacturing, and consumer products. %TC:ignore 
\revDel{For now, potentially the best methods of watermarking AI tools can come about from due diligence and effort on the part of the creators \cite{metfaces}.}
%TC:endignore

\subsection{Enriching creative workflows with text}
The advancements in prompting may push text prompting as a type of interaction into creative tools, even if creative workflows have traditionally not revolved around text. In \nickname, we show the benefit of having a language model scaffold the prompting process. By giving the user fast ways to query and gesture towards what an AI is most likely to understand (as \nickname{} did with the highlighted text options), we enable users to have more opportunities to understand what language may work best with an AI. At the same time, \nickname{} helped users easily reach the design language of their domain, be it robotics or furniture design. In the quantitative survey results, participants felt it was easy to come up with prompts near unanimously for $T_{edit}$ and unanimously for $T_{create}$.

%TC:ignore
\revDel{It is also important to scope out places in the workflow where AI may be of most assistance, whether it be through use cases or through an understanding of what points of entry in a workflow make the most sense. }
%TC:endignore
\revAdd{It is important to understand where in a workflow assistance can be of most use.} Our survey results reflect that \nickname{} produced a slightly more positive experience when it was introduced earlier on in the process. This was corroborated by many participants who said they saw this tool being most helpful in the early stages of design. 
%TC:ignore
\revDel{Ill-placed AI assistance, such as suggesting divergent directions when a model is nearly complete or providing incoherent results that do not ideally respond to prompts could be unproductive for designers. }
%TC:endignore
Well-placed AI assistance, \revDel{however---}such as early stage ideation with GPT-3, trying a text-only prompt to pivot directions, or carefully setting up an image prompt for 3DALL-E to fill in---can be greatly constructive and address painpoints like design fixation that \revAdd{CAD and 3D} designers \revAdd{in general} feel today. Furthermore, if we understand the scope of the tasks we want AI to handle within a workflow, such as having GPT-3 suggest different parts of a model or having DALL-E generate reference images from front, side, and top views, we can better fit general purpose models to their task. %TC:ignore 
\revDel{This is not to imply that we need to finetune the models, but w}
%TC:endignore
We can have stronger checks on the prompt inputs and generation outputs if we understand what is within scope of the task. For example, when P16 wanted a ``flat head'' screwdriver, they were returned results about a medical syndrome---something that could be avoided with content filtering guards checking for relevance to 3D design. AI models may not have to bear the full burden of providing good and ethical answers if we can have multiple checkpoints for propriety.

% Generalizability
\subsection{Generalizability}
The design workflow posed in \nickname{} is generalizable and can easily be used as a blueprint for text-to-image AI integration with different design software. The idea behind surfacing 3D keywords from application related data (as we did with Fusion 360 Screencast data) also introduces ideas for how prompts can be tailored towards the technical vocabulary of a software. The idea of passing in image prompts is also easily extendable to different creative tools, even those outside of the 3D space. For example, graphic editing tools can pass in image prompts based on active layers chosen by a user. Animation software and video editors can send in choice frames for anchored animations and video stylization. A takeaway of this paper is to take advantage of the complex hierarchies that users build up as they design, such as the way \nickname{} takes advantage of the fact that 3D models are generally assemblies of parts. With \nickname,  users could isolate parts and send clean image prompts without the burden of erasing or masking anything themselves.

\subsection{\revAdd{Benefits of Text-to-Image for CAD}}
\revAdd{
Few tools currently explicitly support conceptual CAD \cite{collabcad, frameworkforcad}. \nickname{} supports conceptual CAD not only at the beginning of the design process, but also throughout their workflow, as evidenced by the different usage patterns. It provides visual assets for CAD / product design as well as design knowledge that is otherwise difficult to collect (e.g. standard designs, specific part terminology). These visual assets can be utilized for detailed sketching within CAD, for appearance editing through materials, or for the inspiration of design considerations. 
\nickname{} also presented directions that can solve weaknesses of existing generative tools (\gdnickname) for CAD. By having \nickname{} define shapes and then having the \gdnickname environment optimize them, existing generative tools could better align with what designers visually want, and go beyond physical constraints like loads and forces.}

We demonstrated the efficacy of \nickname{} at supporting a diverse set of potential \revAdd{CAD} end users: mechanical engineers, industrial designers, roboticists, machining specialists, and hobbyist makers. \nickname's interdisciplinary \revDel{nature} \revAdd{design knowledge} is both a strength of AI pretraining as well as the ability of designers to make integrative leaps to meet the AI halfway \cite{stolenelephant}. %TC:ignore
\revDel{Software companies and their customer clients may each have their own domains and internal tools, but the modular nature of \nickname{} facilitates its integration across different settings. } 
%TC:endignore
Additionally, the modular nature of \nickname{} in Fusion 360 demonstrates an idea of separating out AI assistance from traditional non-AI direct manipulation features. Lastly, the text-based nature of the tool and its ready acceptance with designers demonstrates how text interactions can facilitate a low threshold, high ceiling design tool for CAD \cite{myers2000}.

\subsection{Future Work and Limitations}
\revAdd{
A necessary line of future work to make text-to-image AI more usable for CAD will be to integrate it with sketch-based modeling. Sketching is fundamental to CAD and reliant on the creation and manipulation of clean primitives (splines, lines, etc.), and controlling the composition of text-to-image generations based on sketches would be highly useful.}

In terms of limitations, \revAdd{\nickname{}, owing to its implementation in a CAD software, is object-oriented and intended to support CAD product designs (and not 3D art more broadly). It can also occasionally return prompt suggestions that are imperfect or irrelevant to their category (e.g. a ``cylinder'' suggestion could be categorized as both part or design).} There were also times during the study when we experienced technical difficulties. For example, some participants had their DALL-E results cancelled. Moreover, when participants tried to compare the generative design environment with \nickname, the software crashed, so we were unable to directly compare \nickname{} with \gdnickname. However, the existence of \gdnickname, a cloud-based generative design tool \revAdd{for Fusion 360}, shows that there are already CAD designers who utilize generative assistance, and our interviews illustrate that they are open to it improving further. As such, future work can explore how these tools could merge, as \nickname{} has the potential to help with text-based exploration of \gdnickname outcomes. Text-to-3D methods will also be meaningful to explore as they mature in capability of expression, become faster to run at inference, and become more widely available.

Data privacy will also be a key concern in the future. Design know-how and details are the intellectual property of companies and is safeguarded by high-value product industries (i.e. cars). We asked participants to use non-sensitive files, but in the future it will become important to understand how intellectual property passed to AI systems can be protected and not given as free training data. While each prompt needs to be examined for content policy and ethics adherence, there are looming trade-offs to be made in data privacy and AI regulation.

%TC:ignore
\revDel{
The work primarily investigated how well text-to-image tools could be embedded in a workflow where the designs grow increasingly complex. Future work can investigate whether text-to-image AI can provide inspiration that can hold across many different image frames. Rather than producing one image frame of inspiration, we can look at how these models can be embedded into animation pipelines, places where creators are already partially used to automation and computational aids. }

%TC:endignore
\subsection{Broader Impact}
Text-to-image methods have entered the mainstream conversation as a tool that has the potential to impact creative jobs and livelihoods. People have begun to utilize these methods to generate logos, vector illustrations, fashion designs, and so on. This paper is a case study for how generative AI tools can be integrated within the conceptual CAD design stages and how CAD design processes can be augmented rather than automated away. There are ongoing discussions about copyright and existing artist work being leveraged as training data that are rightfully merited. However, we believe that the positive response from participants to \nickname{} illustrates the utility that these tools can present to creatives. Key aspects we think are important for these tools to be successful are that they are narrowed in scope, introduced at early stages of the design process, and still leave room for the creative to exercise their artistic license.

\section{Conclusion}

\nickname introduced text-to-image AI into 3D workflows and was evaluated in an exploratory study with 13 designers. This study elaborated a number of use cases for text-to-image AI from providing reference images to facilitating collaboration to inspiring design considerations. From participant prompts, we observed different types of prompting patterns depending on whether the user engaged with \nickname{} first, last, or throughout their process. Furthermore, we provided measures of prompt complexity across participants and propose a concept for tracking human-AI design history through prompt bibliographies.

\begin{acks}
We thank the AI Lab at Autodesk Research and OpenAI.
% shortened to avoid orphan.
% for their help.
\end{acks}


%TC:ignore
\bibliographystyle{ACM-Reference-Format}
\bibliography{citations2}
%TC:endignore

\end{document}

\section*{Supplemental material for submission:\\ ``3DALL-E: Integrating Text-to-Image AI in 3D Design Workflows''}

\subsection*{Questions for NASA-TLX}
\begin{itemize}
    \item  Effort. How hard did you have to work during this task?
    \item  Mental Demand. How mentally demanding was this task?
    \item Temporal Demand. How hurried or rushed was the pace of your task?
    \item Frustration.  How frustrated were you during the task?
    \item  Performance. How successful were you at accomplishing what you were asked to do?
\end{itemize}

\subsection*{Questions for Creativity Support Index}
\begin{itemize}
\item Goal satisfaction. How much do you agree or disagree: "I was able to find at least one design that satisfied my goal"?
\item Exploration: How much do you agree or disagree: "The system helped me fully explore the space of potential designs"?
\item Enjoyment. How much do you agree or disagree: "I enjoyed working with the system."?
\item Control. How much do you agree or disagree: "I felt like I had control over the generations I was creating"?  
\end{itemize}

\subsection*{Workflow-Specific Questions}
\begin{itemize}
\item Usefulness to Workflow. How much do you agree or disagree: "For this task, 360+DALL-E would be a useful addition to my current 3D modeling workflow".
\item Prompting support. How much do you agree or disagree: "For this task, it was easy for me to come up with new ways to prompt the system"?
\item Feature: GPT3. How much do you agree or disagree: "For this task, GPT3 suggestions were useful for helping me create a prompt".
\item Feature: GPT3. How much do you agree or disagree: "For this task, the color of the GPT3 suggestions were useful to me in suggesting what might work well in a prompt.
\item Feature: Image prompts. How much do you agree or disagree: "For this task, my image prompts were incorporated well into the final generation."

\end{itemize}